\documentclass[prc,aps,groupedaddress,superscriptaddress,twocolumn,nofootinbib,showpacs,showkeys,floatfix,a4paper,10pt]{revtex4-1}

\usepackage{graphicx,colordvi,rotating}
\usepackage{multirow,color}
\usepackage{amsmath,amssymb}
\usepackage[mathscr]{euscript}
\usepackage{cancel}
\usepackage{braket}
\usepackage{cases}
\usepackage{color}
\usepackage{mathtools}
\usepackage{natbib}
\usepackage{ulem}
\usepackage{graphicx}
\usepackage{dcolumn}
\usepackage{bm}
\usepackage{natbib}
 \usepackage{enumerate}
\usepackage[dvipsnames]{xcolor}
\usepackage{ulem}

\usepackage[utf8]{inputenc}
\usepackage{epstopdf}

\usepackage{t1enc}
\usepackage[utf8]{inputenc}

\begin{document}


\title{Scission configuration in the self-consistent calculations with neck constraint}

\author{R. Han}
\email{rui.han@umcs.pl, ORCID: 0000-0002-1687-5157}
\affiliation{Instytut Fizyki, Uniwersytet Marii Curie--Sk\l odowskiej, Lublin, Poland}

\author{M. Warda}
\email{michal.warda@umcs.pl, ORCID: 0000-0001-6723-1020}
\affiliation{Instytut Fizyki, Uniwersytet Marii Curie--Sk\l odowskiej, Lublin, Poland}

\author{A. Zdeb}
\email{anna.zdeb@cea.fr, ORCID: 0000-0001-6965-377X}
\affiliation{CEA, DAM, DIF, F-91297 Arpajon, France}
\affiliation{Université Paris-Saclay, CEA, LMCE, F-91680 Bruyères-le-Châtel, France }

\author{L. M. Robledo}
\email{luis.robledo@uam.es, ORCID: 0000-0002-6061-1319}
\affiliation{Departamento de F\'\i sica Te\'orica and CIAFF, Universidad Aut\'onoma de Madrid, Madrid, Spain}
\affiliation{Center for Computational Simulation,
Universidad Polit\'ecnica de Madrid,
Campus de Montegancedo, Bohadilla del Monte, E-28660 Madrid, Spain
}
\date{\today}

\begin{abstract}
The calculations of the potential energy surface are essential in the theoretical description of the fission process. 
In the constrained self-consistent approach, the smooth evolution of nuclear shape is described from the ground state until a very elongated one with a narrow neck. In all microscopic calculations, the rupture of the neck at scission is associated with a substantial change of nuclear matter density distribution and rapid energy decrease. In this paper, we show that there is no discontinuity of the potential energy surface at scission when multi-constrained calculations are applied with the neck constraint. An early rupture of the neck at lower quadrupole and octupole moments is discussed as competitive with the conventional fission path. We discuss the neck properties in the scission configuration. We find that the neck radius in the asymmetric fission mode cannot decrease below 2 fm, and the nuclear matter density cannot decrease below the saturation density. In the compact fission mode, nuclear density may go down to half of the saturation density before the rupture of the neck.

\end{abstract}

\pacs{}

\keywords{spontaneous fission, potential energy surface, microscopic methods, scission, neck constraint, No-258}
\maketitle


\section{Introduction}

Nuclear fission is a quantum-mechanical process in which a nucleus splits into two similar fragments. It can be understood in terms of tunneling through a potential energy barrier in the space of the parameters describing the shape of the nucleus. Thus, one of the critical elements in the theoretical description of nuclear fission is the determination of the potential energy surface (PES) and the appropriate set of shape parameters, see, e.g., reviews \cite{krappe2012,Schunck2016,10.1088/1361-6471/abab4f}. 
The PES can be obtained from any microscopic nuclear model (Hartree-Fock, Hartree-Fock-Bogolubov (HFB), or covariant density functional) employing constrained self-consistent calculations. These represent efficient and successful methods of describing the fission process, providing a lot of physical information such as fission barrier heights, half-lives, fragment mass asymmetry, etc. \cite{Afanasjev2010,Lu2014,Guzman2014,Guzman2016,Chai_2018,Warda_2015,Warda18}. They are also the basis for more advanced studies of the dynamic of fission.

In microscopic calculations, the PES is usually determined in the space of the quadrupole and octupole moments, describing elongation and left-right asymmetry of the nuclear shape, respectively. More advanced studies also include triaxial shapes or the axial hexadecapole moment used to describe the necking of the nuclear surface. It has to be mentioned that the PES obtained in the constrained self-consistent method may contain discontinuities when treated as a two-dimensional surface \cite{zdeb2021}. This is the unwanted consequence of representing the energy in a two-dimensional grid when it depends in reality on the huge set of parameters characterizing the mean-field wave functions. From now on, we will denote such a set as ``the full space of deformations'' just to distinguish it from the two or three deformation parameters used to characterize the PES. The discontinuities may appear when there are multiple local minima in the full space of deformations, and jumping between them is visible as a sharp ridge on the two-dimensional smooth surface. 

The best-known example of such discontinuity is the so-called {\it scission cliff}. It is the frontier between two types of nuclear shapes: the pre-scission configuration where the whole nucleus is formed by two pre-fragments connected by a neck, and the post-scission configuration where two fragments are a few fm apart. Usually, there is a few MeV difference in energy between these two solutions with the same quadrupole and octupole deformation.
Since the pre- and post-scission configurations are not linked in the usual space of deformation parameters, this causes difficulties in the description of the fission process associated with the determination of {\it the scission configuration},
intuitively defined as a shape of the system when the mother nucleus splits into two fragments. This configuration plays a crucial role in the description of the fragment mass asymmetry. After scission, it is assumed that the nucleons are well localized in each fragment, and therefore the transfer of nucleons between fragments is impossible. The number of protons and neutrons in each of them must be fixed.

Various models have been developed in the past to determine the scission configuration. For instance, in the liquid drop model, it is defined as the configuration where two fragments of uniform density touch each other at precisely one common point \cite{krappe2012}. 

In more realistic models of the nucleus with surface diffuseness, the scission configuration can not be so easily defined since the tails of the nuclear densities of the fragments always overlap near scission. In the droplet model and various macroscopic-microscopic approaches with leptodermous matter distribution, the natural extrapolation of the liquid drop definition of touching fragments leads to describing the scission configuration as the one corresponding to the half-density contour of the two fragments touching at one point.

The scission configuration may also be inferred from the analysis of the evolution of the single-particle energies with the deformation parameters. In this approach, it is defined as the point where the potential energy barrier between the nascent fragments reaches the level of the Fermi surface. This condition guarantees the blocking of nucleons transfer between fragments \cite{krappe2012}.

In the scission-point model, it is assumed that statistical quasi-equilibrium among collective degrees of freedom is obtained at scission \cite{wilkins}. In this approach, the scission configuration is reached when the nuclear interaction at scission is at the level of the binding energy of a nucleon. It is achieved when the spheroids representing the fragments are separated by a distance $d=0-2$ fm (a parameter of the model).  It is worth noting that this approach gives a larger inter-nuclear distance than the liquid drop model but ensures that the transfer of nucleons between fragments is forbidden.

In Refs. \cite{Bonneau07a,Bonneau07b}, the authors defined the scission configuration through a condition imposed in the nuclear interaction between fragments. They assumed that the ratio of the nuclear to Coulomb energies of the interaction between fragments should be as small as $\eta=0.1-0.5 \%$, which is also a model parameter.

In the constrained self-consistent calculations, {\it the scission line} on the PES should separate the fission valley from the fusion one. Such configurations are expected to be somewhere on the observed cliff on the energy, but their localization is not well established and depends on the precise definition used. What can be easily determined is the {\it pre-scission line} at the edge of the fission valley. It is created from the set of mesh points in the space of deformation parameters which are one step-size away (e.g., in quadrupole moment) from mesh points for which only two fragments solution can be obtained in the self-consistent procedure, even if the initial wave-function was taken from the pre-scission line. 

In order to investigate the evolution of the fission process and the nuclear configurations at scission, we perform in this paper multi-constrained self-consistent calculations for the nucleus $^{258}$No using the HFB model with a finite range interaction of the Gogny type. This nucleus is selected as one of the typical heavy actinides nuclei. The half-life of $^{258}$No is $t_{1/2}=1.2(2)$ ms and fission is the principal decay channel in this nucleus \cite{audi20170301}. The PES of this nucleus was also discussed in a previous paper \cite{zdeb2021} using the same framework.
	
By using in addition to the quadrupole and octupole moments a constraint in the neck parameters, we will show that the PES of $^{258}$No is a smooth surface in the transition from the pre-scission to the post-scission configurations, and therefore fission is a continuous process in this nucleus when considering the three deformation parameters. The continuity of the energy near the scission configuration makes it possible to explore the evolution of the wave functions from the whole nucleus to the two separated fragments situation. During the calculations, the neck thickness is controlled with precision in the scission region. The constraint on the neck parameters is similar to the one on the hexadecapole moment, which also controls the width of the neck, but it is more sensitive to tiny modifications of the neck thickness at scission.
	
Using the PES on the multidimensional space with the additional neck constraint, we explore possible pre-scission configurations at smaller quadrupole and octupole moments than those at the end of the asymmetric fission path. This unconventional fission pattern will serve as a competitor against the traditional fission path. The properties of the neck near the scission configuration are also discussed in detail.

\section{Theoretical framework \label{THEORY}}

All the parameters involved in our description of fission are obtained within a mean-field framework, including pairing, namely the HFB approximation. In the HFB method, one seeks a Bogoliubov canonical transformation to quasiparticle creation and annihilation operators in such a way that the HFB energy computed with the HFB vacuum is an absolute minimum \cite{Schunck2016}. In order to find the Bogoliubov amplitudes, we use a minimization algorithm based on the gradient method with an approximate second-order derivative \cite{PhysRevC.84.014312}.
The minimization is a constrained one as we have to enforce that the average number of protons and neutrons coincide with the ones of the nucleus under consideration. We also impose additional constraints on the shape parameters used to obtain the PES for the fission process. Those constraints are easily considered by imposing the energy gradient to be orthogonal to the gradient of the constraints. This requires the introduction of Lagrange multipliers for all the different constraints. To be more specific, we minimize the mean value of the Routhian
\begin{equation}
\delta(\left \langle{\Phi}\right | \widehat{H}-\lambda_Z \widehat{Z}- 
\lambda_N \widehat{N}-\sum\limits_{ij} \lambda_{ij} 
\widehat{Q}_{ij}\left|{\Phi}\right>)=0.
\end{equation}
where  $\lambda_N$ and $\lambda_Z$ are the Lagrange parameters fixing the numbers of neutrons $N$ and protons $Z$, while  $\lambda_{ij}$ is the Lagrange parameter associated with the average value of the multipole moment. Axial quadrupole $(Q_{20})$, octupole $(Q_{30})$, and hexadecapole $(Q_{40})$ deformations were considered in this work. For the Hamiltonian, we use the density-dependent phenomenological interaction of Gogny with the parametrization D1S \cite{Berger90}.  The use of this parametrization is well justified first of all because it was fitted to fission barriers in the actinides and last but not least by a large variety of phenomena described by the parametrization in many different scenarios across the nuclear chart \cite{Peru2014,Robledo2019}.
The quasiparticle operators are expanded in a harmonic oscillator basis which oscillator lengths are optimized for each specific set of collective deformations. The calculations were performed in the axial regime in the deformed oscillator basis with  $N_\perp=15$ shells in the perpendicular direction and $N_z=22$ shells in the $z$ direction. This basis is well adapted for fission studies as it pays more attention to the axis where the matter distribution is most elongated. Such a large number of shells implies a high computational cost and also requires the use of special formulas for the matrix elements of the Gogny interaction to avoid numerical uncertainties \cite{EGIDO199713}.
Beyond mean-field two-body kinetic energy correction and the rotational energy correction \cite{Robledo2019} are included in the results. 

The multipole moments are defined in the traditional way in terms of the associated Legendre polynomials $P_{lm}$ 
\begin{equation}
Q_{lm}= r^l P_{lm}[\cos (\Theta)] \;,
\end{equation}

The numerical computation uses the axially symmetric HFB solver {\it HFBaxial}.
The program starts from an initial wave function and minimizes the energy for a given set of constraints using the gradient method mentioned above. 
Once the minimum is found, the shape of the matter distribution of the nucleus is such that it fulfills the conditions imposed by the constraints. 
The initial configuration is usually taken from a neighboring, previously computed mesh point to speed up the calculations. In this way, the starting point is likely to be close to the minimum, and therefore the initial wave function only requires modest modifications, and consequently, a reduced amount of gradient steps,  to solve the HFB equations.
Because the gradient method is only guaranteed to land in a local minimum, changing the starting wave function might lead to a different (and in some case deeper) minimum as it will be discussed below. In those cases, we say that the HFB solution depends on the starting point. When we jump from one minimum to another as the values of the constraints are varied, we may eventually witness the appearance of jumps in the energy when plotted against the constraint parameters. This situation can be avoided in some cases by sweeping the PES backward in the constraint parameters.

\begin{figure}
\includegraphics[width=0.7\columnwidth, angle=0]{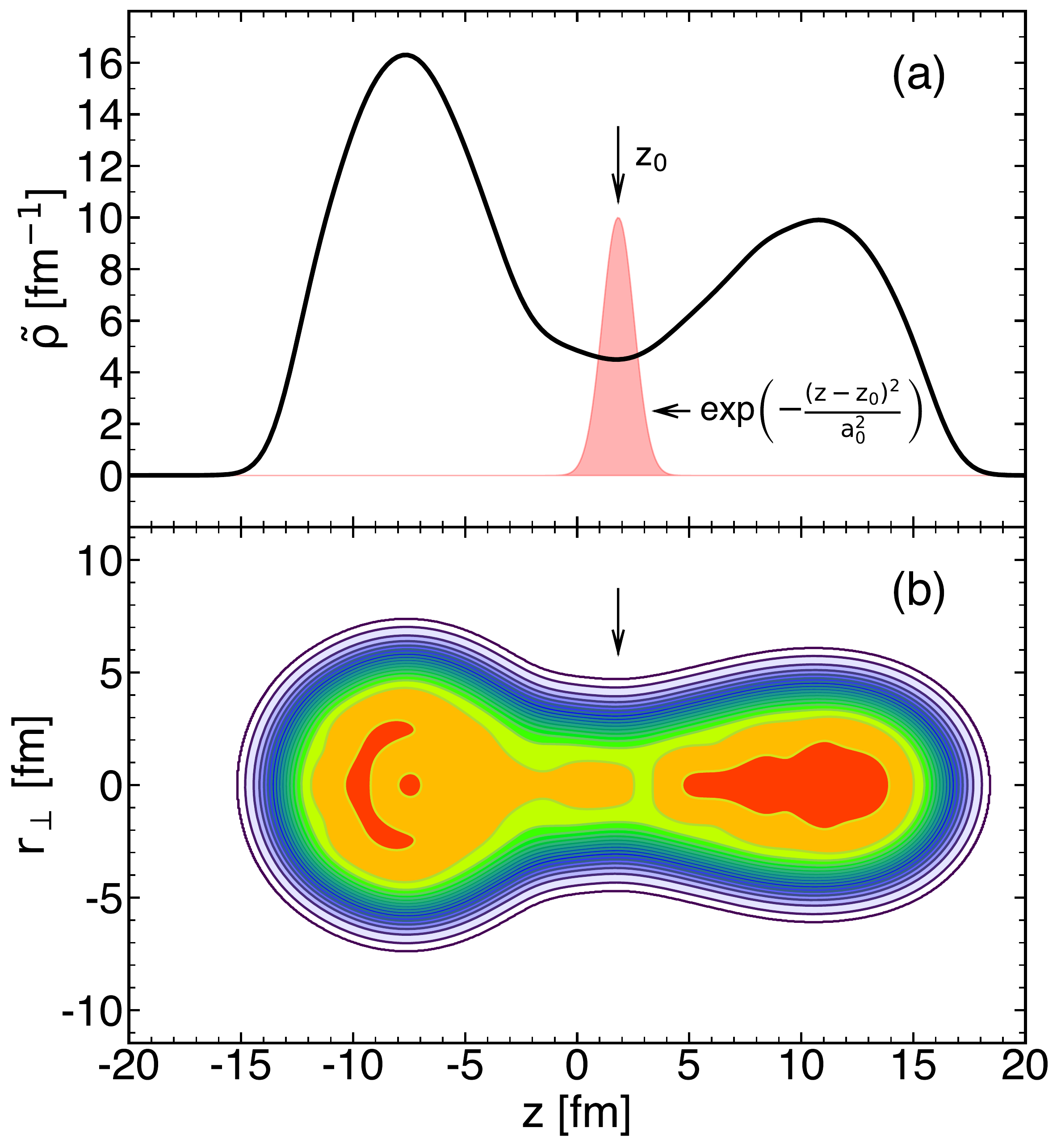}
\caption{(a) Linear density $\tilde{\rho}(z)$ (Eq. \ref{linear_dens}) corresponding to the nuclear density distribution is presented in panel (b). The Gaussian function of the neck constraint (Eq. \ref{qn}) (in arbitrary units) located at the minimum of $\tilde{\rho}(z)$ is also plotted with red line to visualize meaning of the neck parameter $Q_N$ (Eq. \ref{qnpar}).
\label{neckcontraint}}
\end{figure}

\begin{figure}
\includegraphics[width=0.99\columnwidth, angle=0]{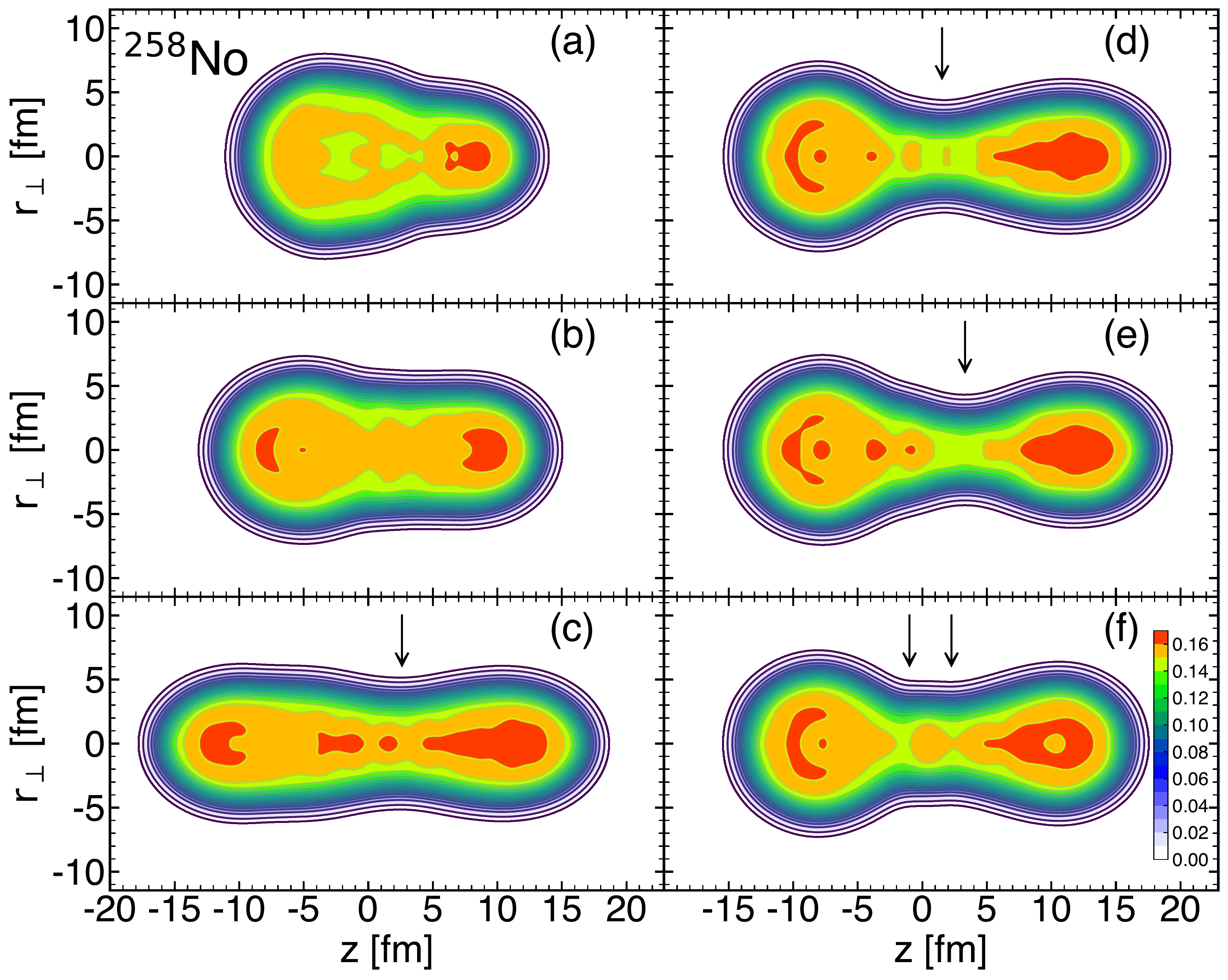}
\caption{Contour plots of various nuclear density distributions are shown here to illustrate the cases where neck position is not well defined. (a) No reduction of linear density, (b) neck as thick as fragment, (c) symmetric elongated shape, (d) and (e) rapid change of $z_0$ (indicated by arrows) in long cylindrical neck. Plots are made for $Q_{20}=200$ b and we find in (c) $z_0=1.60$ fm for $Q_{30}=70$ b$^{3/2}$ and in (d) $z_0=3.21$ fm for $Q_{30}=80$ b$^{3/2}$, (f) double minimum of linear density. See text for details.}
\label{contour}
\end{figure}

In addition to the above multipole constraints we use in this paper, the neck constraint $Q_{N}$ that provides much more precise control of the scission region. The neck constraint is defined through the mean value of the neck operator~\cite{Berger90}:
\begin{equation}
\hat{Q}_N=\exp\left[{-\frac{(z-z_0)^2}{a_0^2}}\right]
\label{qn}
\end{equation}
that depends on two parameters, the position of the neck along the $z$ direction, $z_{0}$ and a width parameter $a_{0}$. 

Roughly speaking, $Q_{N}$ counts the number of nucleons in a slice of nuclear matter centered around $z_{0}$. The width of such slice is approximately equal to the $a_0$ parameter multiplied by 1.7.
We have taken $a_0=1$ fm in all our calculations. With this relatively small value, we make sure that variations of linear density far from $z_0$ (the influence of the neck curvature) do not affect in a relevant way the value of $Q_N$.

The parameter $Q_N$ associated with such constraint gives a good measure of the necking in the final fission phase, far beyond the saddle. Please note that the neck parameter can be shown in terms of one-dimensional density along the $z$ axis
\begin{equation}
{Q}_N=\int_{-\infty}^\infty \tilde{\rho}(z)\hat{Q}_N\; dz \;,
\label{qnpar}
\end{equation}
where the ``linear density'' is given by
\begin{equation}
\tilde{\rho}(z)=\int_0^\infty \int_0^{2\pi} \rho(r,z,\varphi) \;d\varphi dr\;.
\label{linear_dens}
\end{equation}

In Fig. \ref{neckcontraint} we have plotted the linear density for one of the configurations along the asymmetric fission path of $^{258}$No. To visualize the meaning of the neck parameter, we have also plotted the Gaussian function of the neck constraint with $a_0=1$ fm localized at the minimum of the linear density.

The neck position $z_{0}$ should be located at the local minimum of the linear density.
In some cases, however, $z_0$ is not well defined by the minimum of $\tilde{\rho}(z)$, which can even provide misleading information. 
First, it may happen far from the scission at low quadrupole deformations, where the neck has not been well established. In this case, one cannot determine the minimum of $\tilde{\rho}(z)$ (Fig. \ref{contour}a), or the neck is of the same size as the fragments (Fig. \ref{contour}b).
Second, in the so-called symmetric elongated mode \cite{staszczak09,Staszczak13} (Fig. \ref{contour}c), a nucleus takes an almost cylindrical shape, without fragments forming spheroids connected by a neck. It is hard to consider the reduction of the thickness of a nucleus as the formation of a neck.  
Third, at high quadrupole moments the neck usually covers a long region. The linear density $\tilde{\rho}(z)$ can take a minimum at different $z_0$ values for similar deformations. In such cases, the value of $z_0$ does not provide useful information about the localization of the neck, see Fig. \ref{contour}d and e.  Finally, the shape with a double minimum of $\tilde{\rho}(z)$ in the neck region is presented in Fig. \ref{contour}f.

There are two options to determine the value of the $z_0$ parameter. The first way is to calculate it dynamically in each iteration of the self-consistent procedure. It guarantees that $z_0$ will be adjusted to the shape that minimizes the energy. 
The alternative option is to fix the value of $z_0$. In this way, $z_0$ is selected in the neck region but not necessary to be strictly constrained at the position of the thinnest point. This choice allows for controlling the neck region when $z_0$ is taken within a distance not larger than 1 fm from the minimum of the linear density (\ref{linear_dens}). However, both options have limitations. First, the self-consistent calculations cannot be converged with dynamical $z_0$ due to the possible double minimum at neck region, when $Q_N$ is larger than the values obtained from calculations without the neck constraint in the fission valley. On the other hand, when $Q_N$ is very small, the fixed $z_0$ may not allow the precise controlling of the nuclear shape, since the neck position may vary for the same quadrupole and octupole deformation parameters.
Several self-consistent calculations with the neck constraint have been made in previous studies \cite{War11,Warda_2015,wardastaszczak,Warda18,War02,Berger90,schunk,Sadhukhan17,Verriere19}.
They have shown the efficiency of this method in the description of the scission of a nucleus.

One of the goals of this paper is to show that it is possible to obtain a continuous PES  at scission when the neck constraint is applied. In order to check the continuity of the PES we use the {\it density distance} $D_{\rho\rho'}$~\cite{DUBRAY2012} defined by
\begin{equation}
D_{\rho\rho'}= \int|\rho({\bf r})-\rho'({\bf r})| d \bf r\;,
\label{dens_dist}
\end{equation}
where $\rho({\bf r})$ and $\rho'({\bf r})$ are nuclear density distributions calculated for two neighboring points of the mesh of collective variables. This quantity should remain roughly constant and small when there is a smooth evolution of the nuclear shape, while an individual and distinct peak with a large $D_{\rho\rho'}$ value indicates a rapid change of the nuclear density. Such peaks indicate points or regions of discontinuity of the surface \cite{zdeb2021}.

\section{Results}
 
\subsection{The PES of $^{258}$No \label{subsPES} }

\begin{figure}
\includegraphics[width=0.99\columnwidth, angle=0]{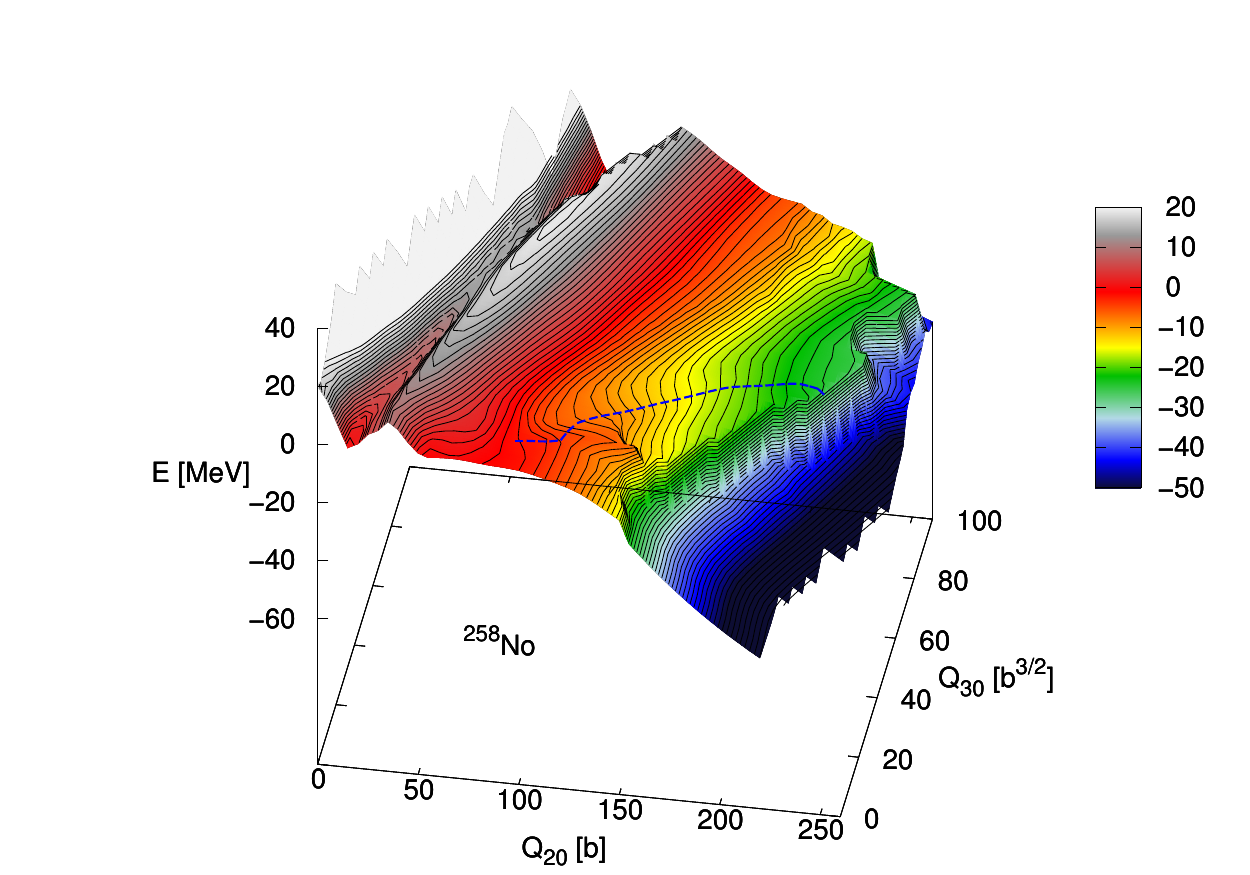}
\caption{The PES of $^{258}$No in a $Q_{20} - Q_{30}$ plane. Constant energy lines are plotted every 1~MeV. Asymmetric fission path is plotted with blue dashed line.\label{PEScfno}}  
\end{figure}

To carry out our investigation on the evolution of the fission process and the nuclear configurations at scission, $^{258}$No is chosen as one of the typical heavy actinides nuclei. The half-life of $^{258}$No is short as $t_{1/2}=1.2(2)$ ms and fission is the principal decay channel in it. The PES landscape of $^{258}$No is plotted in Fig. \ref{PEScfno} as a function of the quadrupole and octupole deformation parameters, which roughly determine the elongation and reflection asymmetry of the nuclear density, respectively. The quadrupole deformed ground state is separated from scission by a fission barrier that goes through reflection symmetric nuclear shapes. Triaxial deformation observed at the first barrier \cite{zdeb2021} is not discussed in this study. At large elongations (from $Q_{20}=100$ b), a broad reflection asymmetric valley opens up.  A small valley of compact fission is located in the vicinity of the $Q_{30}=0$ b$^{3/2}$ line with nearly reflection-symmetric shapes. These distinct regions of the PES are commonly referred to as {\it fission valleys}. In these regions, the nucleus is still a compound system, despite being quite elongated for high quadrupole moments. 
The fission valleys terminate with a quite sudden few-MeV fall in energy at the end of the fission path. The rest of the PES called {\it fusion valley} is created from the solutions of the HFB equations containing two well-separated fragments. The two parts of the PES are easy to distinguish as there is a few MeV difference between them. In the 3D visualization of the PES map as shown in Fig. \ref{PEScfno}, it looks like a vertical scission cliff, which has been mentioned in the Introduction.

In the following, we would like to show that the sudden change of the energy at scission is due to the visualization of the energy projected on a two-dimensional map, which is a commonly used procedure in the description of the PES. By applying the neck constraint together with the other two deformation parameters ($Q_{20}$ and $Q_{30}$), it is possible to connect the pre- and post- scission configurations continuously. In such way, we overcome this common problem showing up in the description of scission in all self-consistent calculations.

\subsection{Neck thickness}

\begin{figure}
\includegraphics[width=0.7\columnwidth, angle=270]{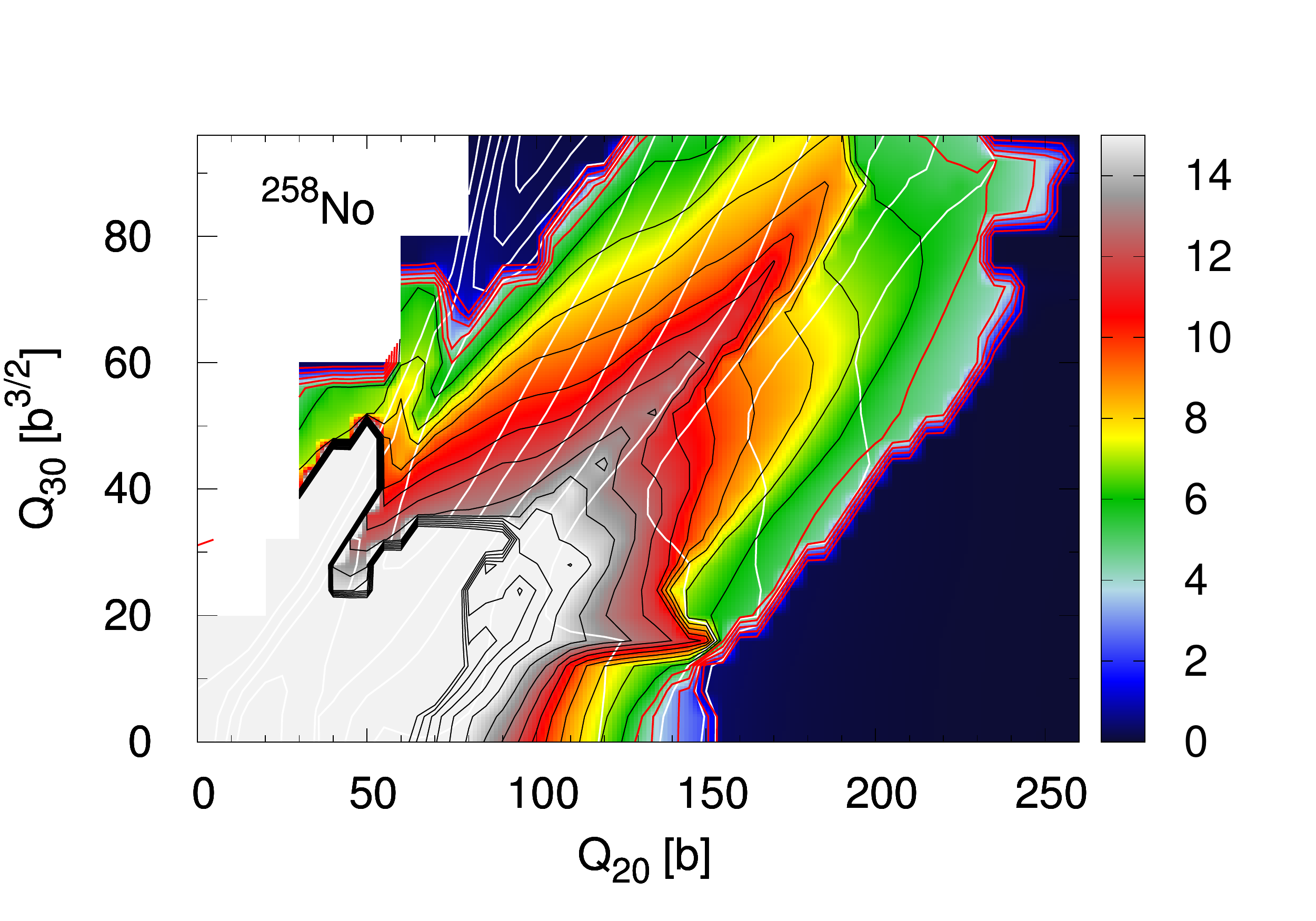}
\caption{Neck parameter $Q_N$ in a deformation space $Q_{20}, Q_{30}$ of  $^{258}$No. Isolines are plotted from $Q_N= 1$ to 5 in red and from 6 on in black with $\Delta Q_N=1$. Constant energy lines are plotted with white lines every 5 MeV for a better identification of regions of the PES\label{qn_cfno1}.}
\end{figure}

\begin{figure*}
\includegraphics[angle=270,width=1.9\columnwidth]{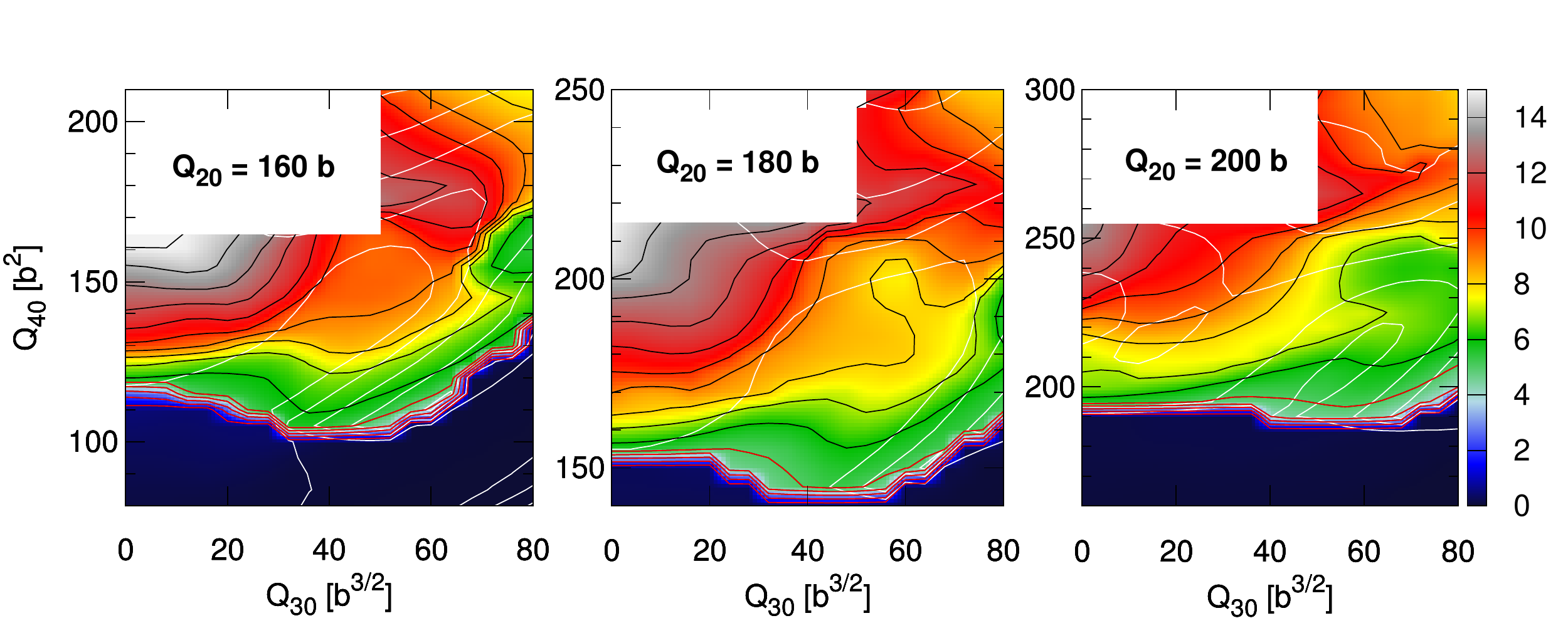}
\caption{Neck parameter $Q_N$ values in a $Q_{30} - Q_{40}$ plane for various values of $Q_{20} $ in $^{258}$No.  The color scale and isolines are the same as in Fig. \ref{qn_cfno1}. 
\label{qn_cfno3}}
\end{figure*}

In order to further investigate the scission configruation, we start our study by analyzing the neck parameter $Q_N$ values obtained consistently for different configurations in the PES on $Q_{20}-Q_{30}$ surface.

In Fig. \ref{qn_cfno1} we plot the neck parameter $Q_N$ defined by Eq. (\ref{qn}) as a function of the $Q_{20}$ and  $Q_{30}$ constrained multipole moments for $^{258}$No. In the definition of $Q_N$, the parameter $z_0$ is located at the minimum of the neck thickness. The other parameter $a_0$, which is related to the width of the neck constraint, is taken as a reasonable value $a_0=1$ fm (Fig. \ref{neckcontraint}).
As shown in Fig. \ref{qn_cfno1}, in the region of small quadrupole deformations and up to the first barrier, the neck parameter is very large or cannot be determined due to the absence of a well-defined neck region. 
In the region beyond the first barrier, the $Q_N$ surface presents the intuitive correlation between the increasing elongation and the decreasing neck thickness, which is commonly used to understand the nuclear shape evolution in fission. 
Finally, the neck thickness drops down to a small value in the scission region, indicating splitting the whole nucleus into two fragments. Note that the $Q_N$  value never goes down to zero since the tails of the nucleons' wave functions never vanish entirely in the space, and the nuclear matter density is positive everywhere. 

In Fig. \ref{qn_cfno1} we can see a slightly different behavior of the neck parameter between the compact fission mode (below $Q_{30}=15$ b$^{3/2}$) and the asymmetric fission mode. The shape evolution along the symmetric fission path leads to a compact configuration with two identical spherical pre-fragments connected by a short neck, while in the asymmetric fission valley, the neck region is much longer and connects the two pre-fragments with distinct deformations. These two cases should be considered separately.

An interesting finding in our study is the value of the smallest neck thickness can be obtained in the whole nucleus. The limiting value of the neck parameter for the pre-scission shape is around $Q_N=4 -5$. It stays roughly constant along the entire pre-scission line in the asymmetric fission valley. An equivalent quantity used to characterize the properties of the neck is the so-called neck radius. Assuming a uniform and sharp cylindrical density distribution in the neck region, the neck radius is defined as the radius of the cylinder of length $\sqrt{\pi}a_0$,
\begin{equation}
r_{\mathrm{NECK}}=\sqrt{{Q_N}/({\rho_0 \pi^{3/2}a_0})}=1.059\sqrt{Q_N}\;,
\end{equation}
with $a_0 = 1$ fm and the standard value for the nuclear density $\rho_0=0.16$ fm$^{-3}$. We obtain $r_{\mathrm{NECK}}= 2.1$ fm, 2.4 fm, and 2.6 fm for neck values $Q_N=4$, 5 and 6, respectively.  
In the self-consistent calculations, the limiting value of the neck radius is closer to the equivalent rms radius \cite{hasse,centelles10} of a deuteron $r=2.76$ fm ($r_\mathrm{RMS}^{\mathrm{charge}}=2.1421(88)$ fm \cite{iaea}) or an $\alpha$-particle $r=2.13$ fm ($r_\mathrm{RMS}^{\mathrm{charge}}=1.6755(28)$ fm \cite{iaea}) than to the one of a proton $r=1.12$ fm ($r_\mathrm{RMS}^{\mathrm{charge}}=0.8783(86)$ fm \cite{iaea}). We can conclude that the nuclear matter distribution cannot take the form of a very thin neck with a layout similar to the one of a chain of nucleons. The effective nuclear interaction among nucleons is attractive at short distances, and when the repulsive Coulomb force between fragments gets stronger, the neck just breaks up instead of creating an extremely thin structure in the neck.

Similar conclusions about the smallest possible neck thickness were drawn from macroscopic-microscopic calculations. It was found that the limiting value of the neck thickness is $(0.25-0.30) R_0$, which corresponds to $1.9-2.3$ fm \cite{ivanyuk09,ivanyuk12}. The droplet model with a random walk approach for dynamical calculations also indicates 2 fm as a critical neck radius restricting the evolution of mass asymmetry \citep{randrup11}.

In the symmetric fission valley, the neck thickness parameter goes down to $Q_N=2$ before rupturing of the neck. Here we have got so-called compact fission with two spherical fragments. This issue will be discussed in detail in the following subsections.

For a more detailed study of the neck parameter at scission, we performed triple constrained calculations where quadrupole, octupole, and hexadecapole moments were simultaneously considered. As it was shown in Ref. \cite{zdeb2021}, such procedure provides much more details concerning the evolution of nuclear shapes in fission. Following the idea presented in  Refs. \cite{zdeb2021,Tsekhanovich2019}, Fig. \ref{qn_cfno3} shows the neck parameter in the  three-dimensional space by plotting a set of slices for fixed $Q_{20}$ as a function of $Q_{30}$ and $Q_{40}$.
As expected, decreasing the hexadecapole moment reduces the value of the neck. Exceptions to this rule are observed for very exotic shapes with high excitation energies, as it can be seen in the upper right corner of each panel. In addition, we can clearly see in this figure that, unfortunately, $Q_{40}$ fails in describing the scission configuration accurately. A slight reduction of the $Q_{40}$ value causes the rapid decrease of $Q_N$, pointing that the scission region has been reached. Again, we can see that the limiting value of the fission valley is around $Q_N=5$. It confirms the earlier findings that the neck constraint has to replace the hexadecapole moment in the scission region for a more precise description of the scission process.

\subsection{Neck constraint calculations}

\begin{figure*}
\includegraphics[width=1.9\columnwidth, angle=0]{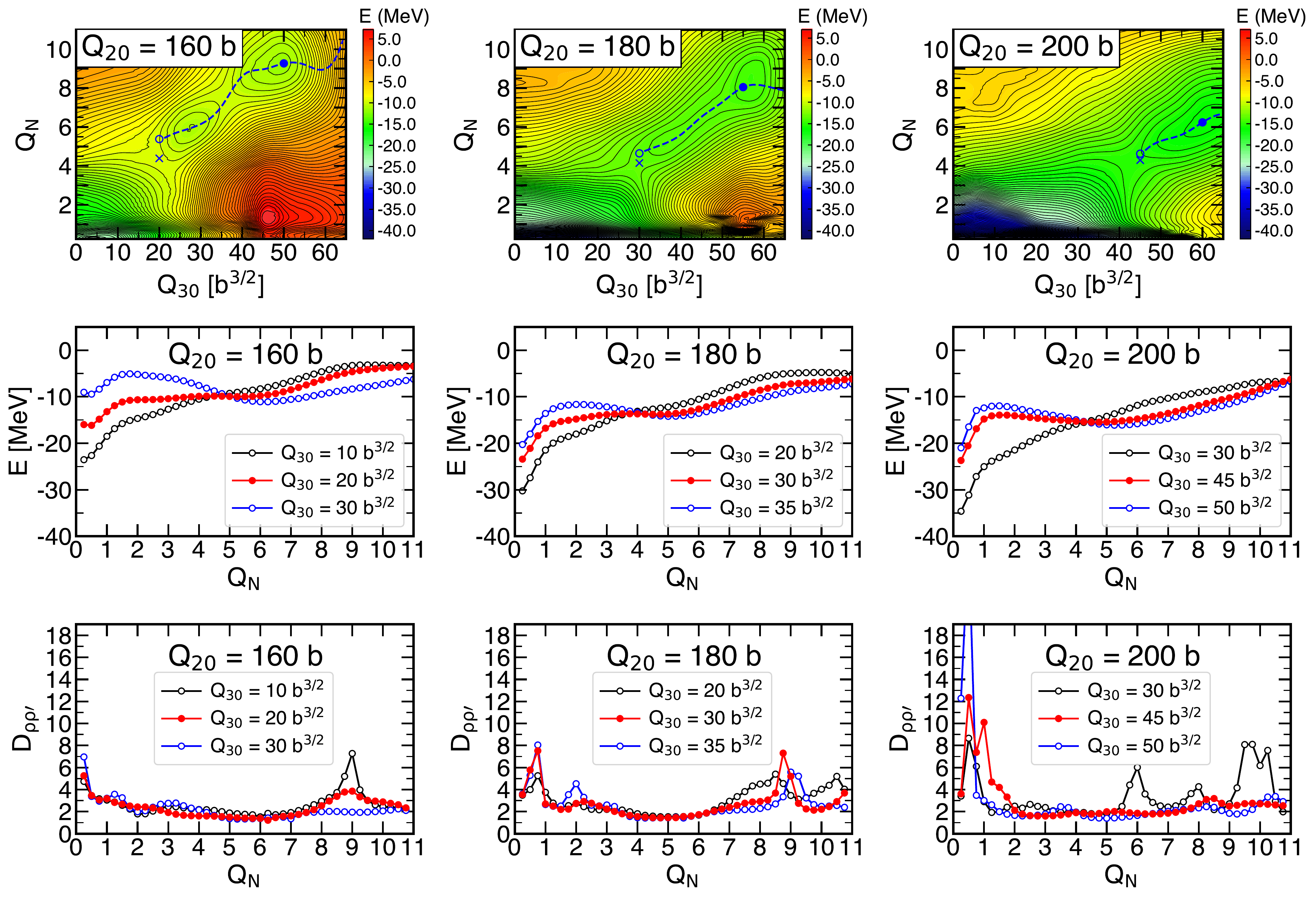}

\caption{Top row: The PES of $^{258}$No  for fixed $Q_{20}=160$ b (left), 180 (center) and 200 b (right) as a function of $Q_{30}$ and the neck parameter $Q_N$. See details in the text. Middle row: Energy profiles for selected $Q_{30}$ values from the upper plots. Bottom row: Density distance $D_{\rho\rho'}$ along lines from the middle row.
\label{qn_q3}}
\end{figure*}

\begin{figure}
\includegraphics[width=1.0\columnwidth, angle=0]{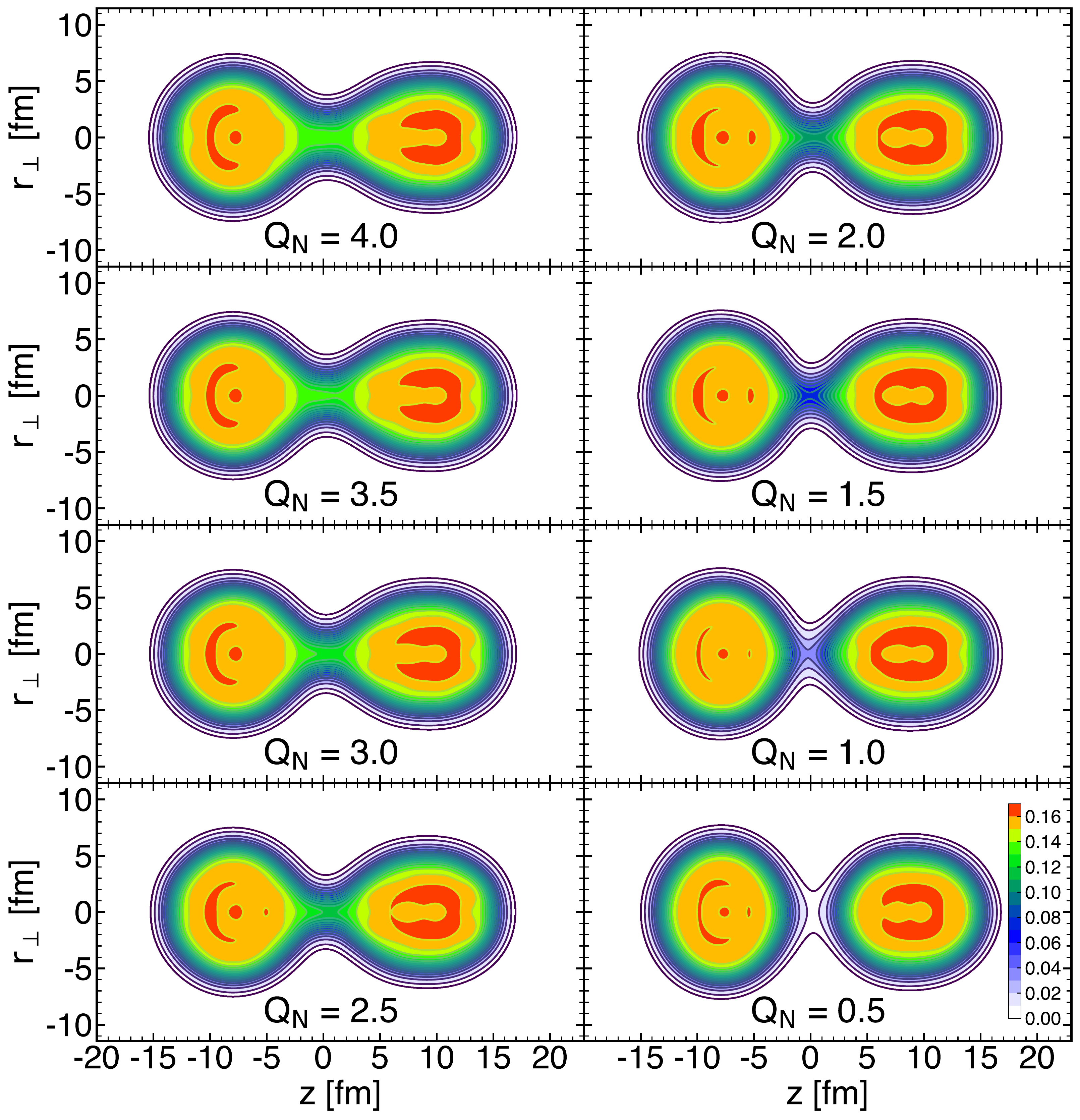}
\caption{The series of nuclear matter density distribution in the region of scission of $^{258}$No for fixed $Q_{20}=180$ b and $Q_{30}=30$ b$^{3/2}$ with various values of $Q_N$ constraint. In this case, $Q_N= 4$ corresponds to the pre-scission configuration (saddle point at Fig. \ref{qn_q3}).
\label{rupture}}
\end{figure}

\begin{figure*}
\includegraphics[width=1.9\columnwidth, angle=0]{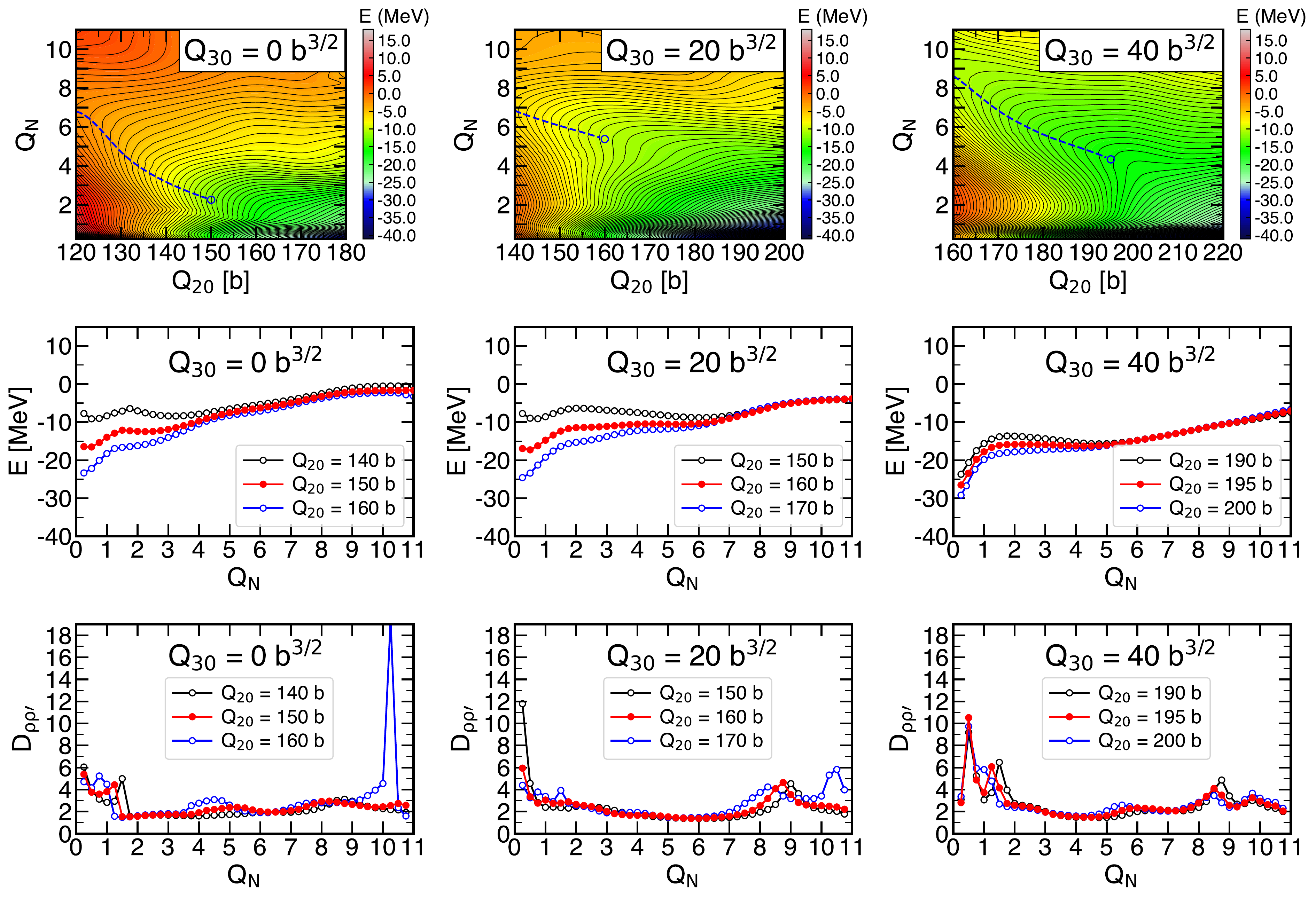}
\caption{The same as in Fig. \ref{qn_q3} but for fixed octupole moments $Q_{30}=0, 20$, and 40  b$^{3/2}$ and as a function of $Q_{20}$ and the neck parameter $Q_N$. 
\label{q2qn}}
\end{figure*}

\begin{figure}
\includegraphics[width=1.0\columnwidth, angle=0]{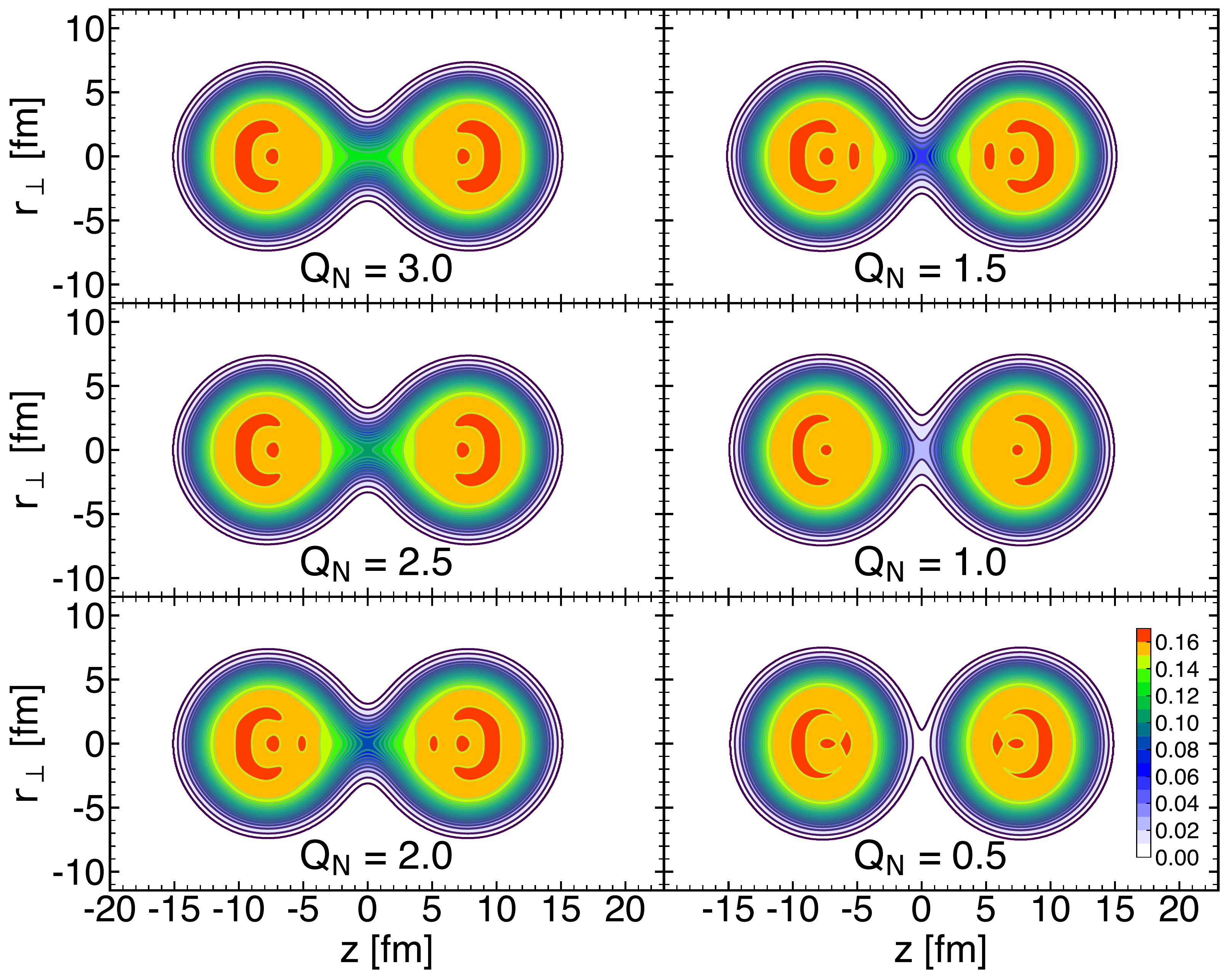}
\caption{The density matter distributions in the region of scission of $^{258}$No  for fixed $Q_{20}=150$ b and $Q_{30}=0$  b$^{3/2}$ (compact fission mode) with various values of the $Q_N$ constraint. With the above values for the multipole moments, $Q_N= 2$ corresponds to the pre-scission configuration (saddle point at Fig. \ref{qn_q3}).
\label{rupturesym}}
\end{figure}

In the top panels of Fig. \ref{qn_q3} we have plotted the PES of $^{258}$No calculated with triple constraints, $Q_{20}$, $Q_{30}$ and $Q_N$. Note that the width parameter $a_0$ in neck constraint is always taken as $a_0=1$ fm in the following calculations. In the surface, the lower part ($Q_N$ smaller than the values obtained consistently from the calculations without neck constraint) is calculated with dynamically adjusted $z_0$, while the upper part (roughly above blue dashed line) with fixed $z_0=0.5$ fm. We obtain surfaces similar to those calculated with hexadecapole constraint $Q_{40}$ (see Fig. 5 of Ref. \cite{zdeb2021}), but with a much better visible structure of the scission region. In the central region of each panel, we can see a clear asymmetric scission valley with a fission path marked by a blue dashed line that corresponds precisely to the data at the $Q_{20}-Q_{30}$ two-dimensional surface in Fig. \ref{PEScfno}. 
The minimum of the valley marked by a full dot is an element of the fission path from Fig. \ref{PEScfno}. In the upper left side of each top panel, there is another shallow valley corresponding to the symmetric elongated fission mode. It is much higher in energy than the asymmetric valley. At the bottom part of these panels, along $Q_N\approx0$, the post-scission configuration is obtained. It is a part of the fusion valley.

Each configuration from a fission valley is separated from a fusion valley by an energy barrier at $Q_N\approx2$. Decreasing neck thickness leads to an increase of energy even over 10 MeV. This explains why the nucleus prefers staying in the fission valley at a relatively low quadrupole moment rather than splitting at the same deformations.

The asymmetric fission valley ends with a saddle, marked with an x-cross in Fig. \ref{qn_q3}, which separates it from the valley with post-scission shapes in the minimum. This saddle has the same deformation parameters as the pre-scission point, marked with an open circle. The minor difference between the localization of these two points is an outcome of using constraints that are not orthogonal to the steepest descent line (see p. 269 of ref. \cite{rin80}). Thus we get a clear interpretation of the previously defined pre-scission line. It corresponds to a set of saddle points in the $Q_N-Q_{30}$ maps at fixed $Q_{20}$ that separate the fission valley from the fusion one.

More intriguingly, we notice that in Fig. \ref{qn_q3}, the minima (full dots) correspond to a series of points along the fission path on the $Q_{20}-Q_{30}$ map (Fig. \ref{PEScfno}). The minimum (full dot) and the pre-scission point (open circle) are connected by a smooth path, and the energy difference between them can be easily overcome with a small value of $\sim$1MeV. This allows the unconventional early rupture of the neck at lower quadrupole (octupole) moment instead of the traditional endpoint of the fission path. Note that this rupture only occurs at the pre-scission point (open circle). Such  fission pattern will surely serve as an important competitor against the traditional fission path. Therefore, the distribution of matter at this saddle point is essential for determining the fission fragments' mass asymmetry.

Considering the fact that the saddle corresponds to a pre-scission configuration, a question arises whether the pre-scission line can be determined for the other octupole moments in the $Q_{30}-Q_N$ map. According to the evolution of the nuclear density for each configuration, it is found that the pre-scission point can be easily extended to the ridge of the barrier separating the asymmetric fission valley from the fusion valley for the octupole moments larger than the saddle. The ridge lies along $Q_N\approx 2$. 
For octupole moments lower than the saddle, however, a completely different scenario is present.  The asymmetric fission valley smoothly converts into the fusion valley without any ridge or other kind of borderline. There is only a barrier at $Q_N=10$ separating symmetric elongated fission valley from the shapes with a thinner neck.
In this region, a pre-scission line cannot be unequivocally established.

For a deeper view into the evolution of energy at scission, we have plotted the energy profiles for several values of $Q_{30}$ selected from the maps in the top panels as a function of $Q_N$, which are shown in the middle panels of Fig. \ref{qn_q3}. For the lines with octupole moments larger than the saddles (lines with blue open circles), there is an apparent minimum of the asymmetric fission valley and a barrier separating it from the shapes without a neck for each line. For the lines going through the saddle-point (full symbols), minima convert into plateaus. For low-octupole-moment lines, only the slope towards scission can be found.

Another important question is whether the passage from the fission valley to the fusion one is continuous or not. From the maps and the energy profiles in the top and middle panels of Fig. \ref{qn_q3}, it seems to be continuous. However, it has been shown \cite{zdeb2021} that visual continuity of the surface may result from the smoothing procedure of the graphical tools used to prepare plots. A more precise characterization of continuity can be made by using the density distance parameter $D_{\rho\rho'}$ (\ref{dens_dist}). In the bottom panels of Fig. \ref{qn_q3}, we have plotted the density distance parameter as a function of $Q_N$ along the lines presented in the middle panels. 
In general, the density distance parameter stays all the time at a constant level, while there are several regions of slightly higher values of $D_{\rho\rho'}$, especially for $Q_N<2$. The most significant effect is at $Q_2=200$ b and $Q_N=0.5$. In this region, nucleons are almost completely pushed out of the neck region, and the neck constraint can hardly control the shape of the nucleus. The shell effects in each fragment play an important role in the nuclear matter distribution here. The most significant effect of discontinuity is visible in the lower right corner of the PES in the top panels of  Fig \ref{qn_q3} as a cliff on the energy barrier. At these extreme deformations, neck constraint is unable to save the uniform structure of the surface, but this region is unimportant as the potential energy barrier takes huge values there.

In Fig. \ref{rupture} we have plotted the evolution of the shape of nucleus for  $Q_{20}=180$ b and  $Q_{30}=30$ b$^{3/2}$ for different values of $Q_N$ constraint.  The saddle (pre-scission) point is at $Q_N=4$. We can see a continuous transition of the shape from the whole nucleus to two well-separated fragments.

Fig. \ref{q2qn} contains the similar information as Fig. \ref{qn_q3}, but the PES is projected on the other set of variables. Here we have plotted the maps in the coordinates $Q_{20}$ and $Q_N$ for fixed values of $Q_{30}=0, 20$, and 40 b$^{3/2}$. The fission valley goes from the upper left to the lower right corner of each panel. At the low quadrupole moment, rupture of the neck is blocked by the energy barrier. The constant energy lines show the ``backbending'' pattern here. Going to larger elongations, the barrier declines and finally disappears.  Pre-scission point (open circle in Fig. \ref{q2qn}) can be found at the quadrupole moment where constant energy lines are almost vertical in the barrier region.  At larger deformation, no stable solution of the whole nucleus can be found unless it is constrained by the necking operator.

In the middle panels of Fig. \ref{q2qn} energy curves are plotted for selected values of $Q_{20}$ from the maps in the top panels. We can see the disappearance of the barrier between fission and fusion valleys. The values of the density distance parameter as a function of $Q_N$ plotted in the bottom panels confirm the smoothness of the PES in this region. No sharp peaks are visible here.

It is well known that the reflection symmetric fission mode shows several distinctive properties that differentiate it from asymmetric fission. The name ``compact mode'' comes from the facts that both pre-fragments have a spherical shape, and the distance between them is relatively small at scission. The properties of the neck are also different from the cases in the reflection asymmetric mode, as can be seen in the left column of Fig. \ref{q2qn}.  The shape of the PES looks similar to those in the asymmetric mode, but the nucleus achieves a much smaller value of the neck parameter $Q_N=2$ in the compact fission pre-scission configuration. The continuous series of nuclear shapes at scission are plotted in Fig. \ref{rupturesym}. In the pre-scission configuration, the density is much lower than the saturation density.

In this way, we have shown the nature of the scission-line cliff of the energy on the two-dimensional maps of the PES. It is a numerical artifact due to the truncation of the collective space. The PES is continuous in a full space of deformations and remains continuous when a proper and large enough set of constraints is applied. Please note that calculation with two constraints on $Q_N$ and $Q_{20}$ (used in the other papers \cite{wardastaszczak,Warda18}) may not be sufficient to obtain a continuous surface at scission. From the other side, constraints on $Q_N$ and $Q_{30}$ create smooth PES in the specific case of super-asymmetric cluster radioactivity fission \cite{War11}.

\subsection{Density profile at scission}
 
\begin{figure}
\includegraphics[width=0.99\columnwidth, angle=0]{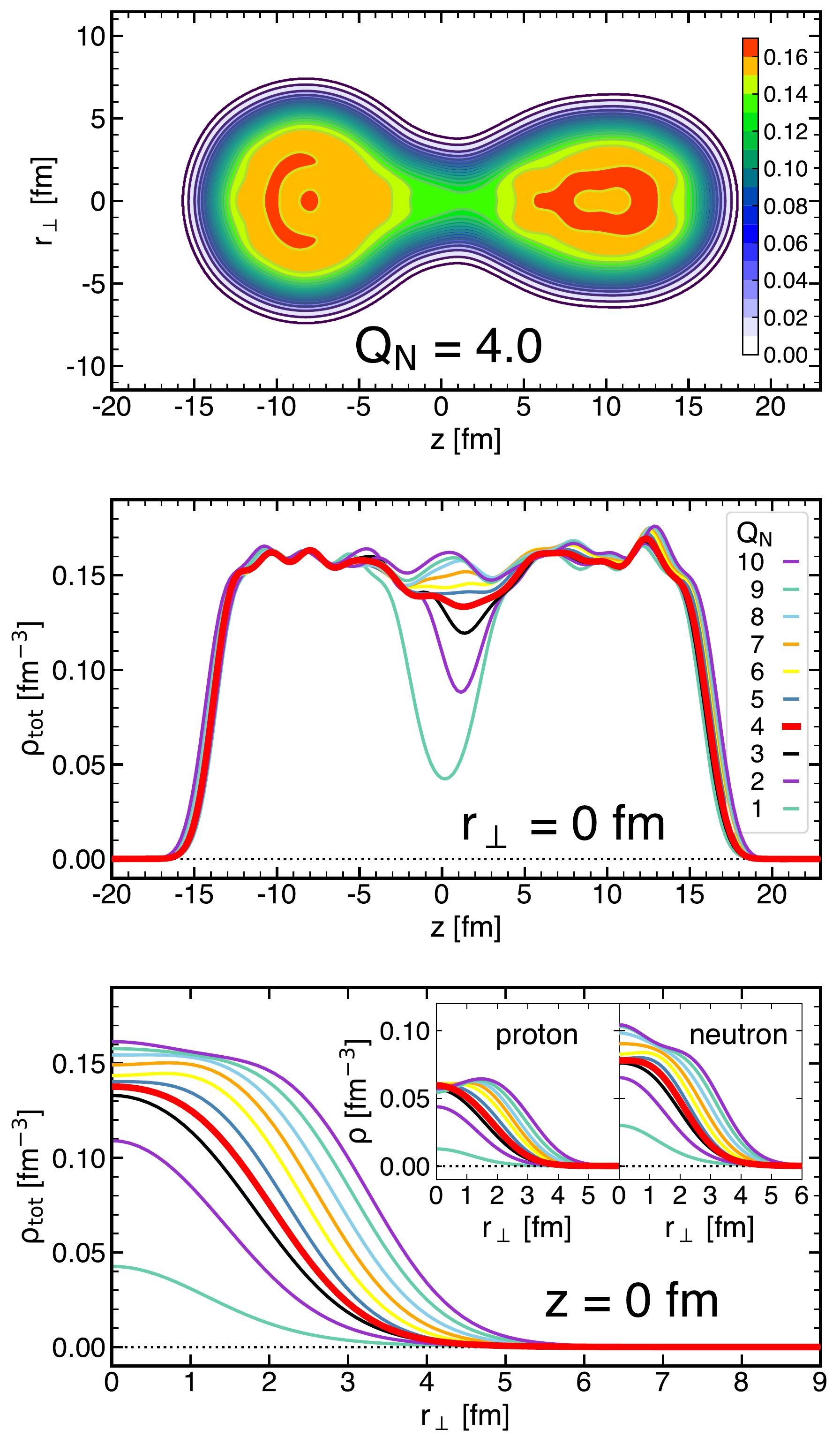}
\caption{(Top) Density distribution of $^{258}$No for fixed $Q_{20}=200$ b, $Q_{30}=45$ b$^{3/2}$, and $Q_N=4$. (Middle) Nuclear matter density profile along the symmetry axis $z$ for the configurations with the same quadruapole and octupole moments as in the top panel but with various values of $Q_N$. (Bottom) The same as in the middle panel but as a function of $r_{\perp}$ at fixed $z_0=0$ fm. Insets present proton matter and neutron matter profiles.
\label{profiles}}
\end{figure}

\begin{figure}
\includegraphics[width=0.99\columnwidth, angle=0]{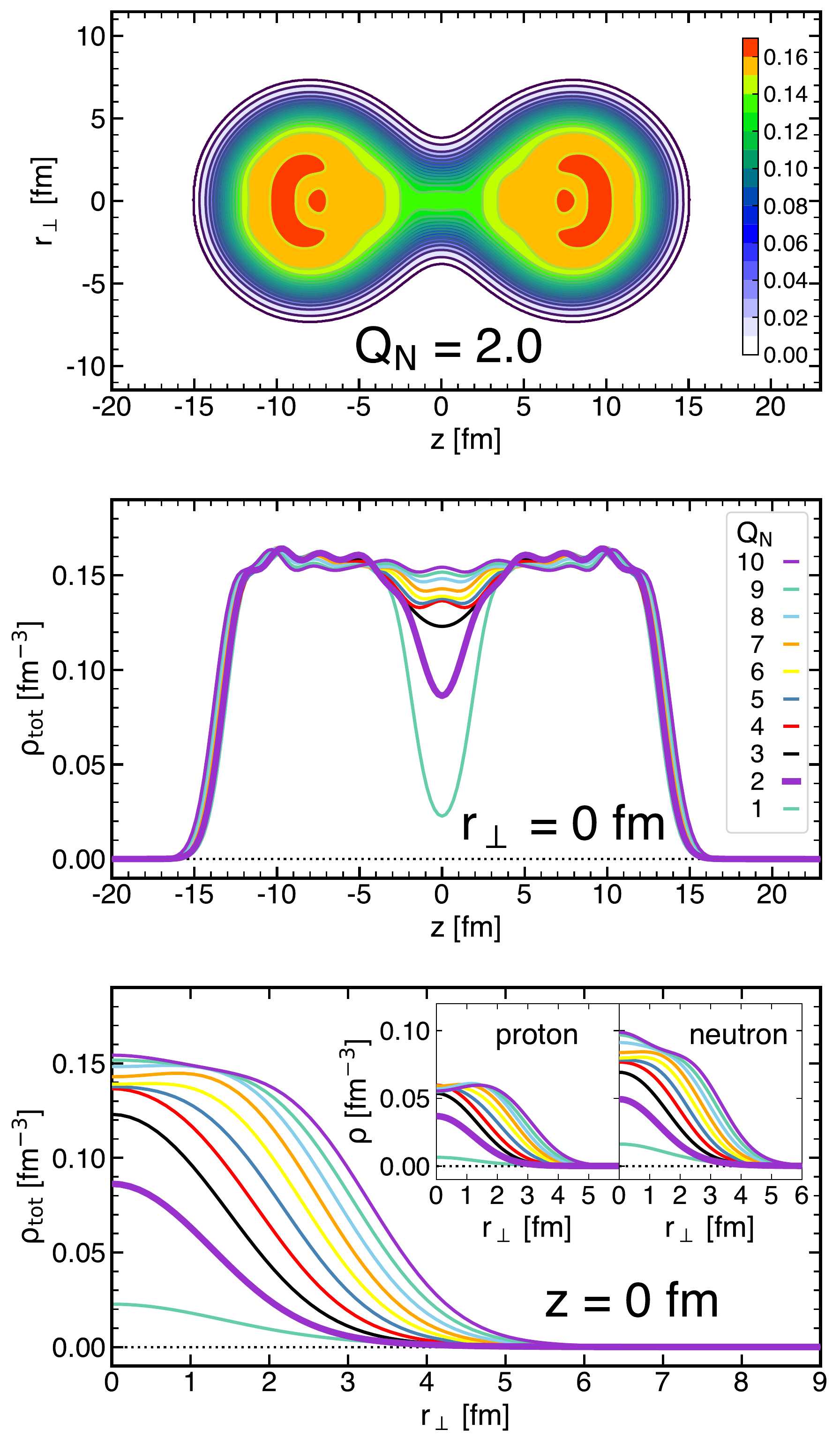}
\caption{The same as in Fig. \ref{profiles}, but for compact mode for fixed $Q_{20}=150$ b, $Q_{30}=0$ b$^{3/2}$. Density profile in top panel is plotted at $Q_N=2$
\label{profilescompact}}
\end{figure}

As mentioned in the previous paragraphs, one of the main advantages of this work is that the scission configuration can be determined precisely. In the top panel of Fig. \ref{profiles}, we have plotted the nuclear matter density distribution of one of the pre-scission configurations at $Q_{20}=200$ b and $Q_{30}=45$ b$^{3/2}$. It is also the saddle point in Fig. \ref{qn_q3} (at $Q_N=4$).  
In the middle and bottom panels of  Fig. \ref{profiles}, we have plotted the density profiles of the nucleus for the same deformation but with various $Q_N$ as a function of $z$ at the symmetry axis (central density), and as a function of radius perpendicular to the symmetry axis at $z=0$ fm, respectively. 
For large values of $Q_N\ge4$, the central density remains at the level of saturation density and goes down at most to 0.13 fm$^{-3}$ at the neck, and fluctuates along the symmetry axis. The density profile as a function of $r_\perp$ takes the shape of the Fermi function, similar to the leptodermous distribution of matter in spherical nuclei. 

In the present neck constraint calculations, the neck thickness is controlled by the $Q_N$ parameter. For $Q_N\ge4$, there are two coexisting effects responsible for the reduction of the neck thickness with $Q_N$ decreasing. First, the half-density radius and the size of the bulk region reduces, while the slope of the surface density remains roughly constant. Second, the saturation density reduces slightly due to the decrease of the neutron's matter density at $r_\perp=0$ fm, while the proton's density remains almost constant at $\rho_{\mathrm{prot}}=0.06$ fm$^{-3}$ for $Q_N\ge 4$, as shown in the insets of Fig. \ref{profiles}. It means that the variations of the central density are strictly related to the shell structure and localization of particular orbitals within the bulk of nuclear matter \cite{warda14}.

For $Q_N<4$, the density profiles show different characters. In the middle panel, it is shown that the density profile at the neck region takes a parabolic shape with a minimum decreasing with the neck parameter. 
In the bottom panel, compared with the lines for $Q_N\ge4$, the plateau of the saturation density vanishes, and the surface density drops starting from the symmetry axis. Therefore, for the post-scission configurations, further reduction of the nuclear matter at neck region is achieved by reducing the central density at the symmetry axis without significantly decreasing the neck radius.

It is worth noting that, from the pre-scission configuration to the post-scission configuration, the neck position $z_0$ substantially shifts from $z_0=2$ fm towards $z_0=0$ fm, corresponding to the evolution from a compound system to two well-separated fragments during the rupture of the nucleus.

The density profiles presented in Fig. \ref{profilescompact} for the compact mode are quite similar to those shown for asymmetric mode. The main difference is that the central density at pre-scission configuration goes down to $0.08$ fm$^{-3}$ in compact mode, which is much lower than the value in asymmetric mode. The reason can be due to the shell structure, as the fragments created in compact mode are close to doubly-magic $^{132}$Sn. These doubly magic isotopes are spherical and very stiff against deformation, making it possible to keep their spherical shape despite the presence of the other nearby fragment. In the compact fission mode, nucleons are unlikely to be localized between pre-fragments. The neck is mainly an outcome of the overlapping tails of the surface diffuseness.
However, in the asymmetric fission mode, the right-hand-side pre-fragment is elongated and octupole deformed. Such nuclei are usually soft against changes of deformation.  It is easy for the fragment to acquire a shape that converges smoothly into a neck in the space between pre-fragments. In this mode, the neck is a few fm long, with the nuclear density remaining roughly constant at the level of 0.14  fm$^{-3}$, which is slightly below the saturation density. This neck must keep its thickness all along its length.

We can conclude that the configurations before scission in the asymmetric mode are determined by the thinnest neck that satisfies the condition of saturation density at the symmetry axis and typical surface diffuseness. The nuclear system with central density at the neck reduced below this value is unstable against the diminishing of the neck. In the symmetric mode, central density may be much lower before rupture of the neck. 

\subsection{Droplet model analysis}

\begin{figure}

\includegraphics[width=0.99\columnwidth, angle=0]{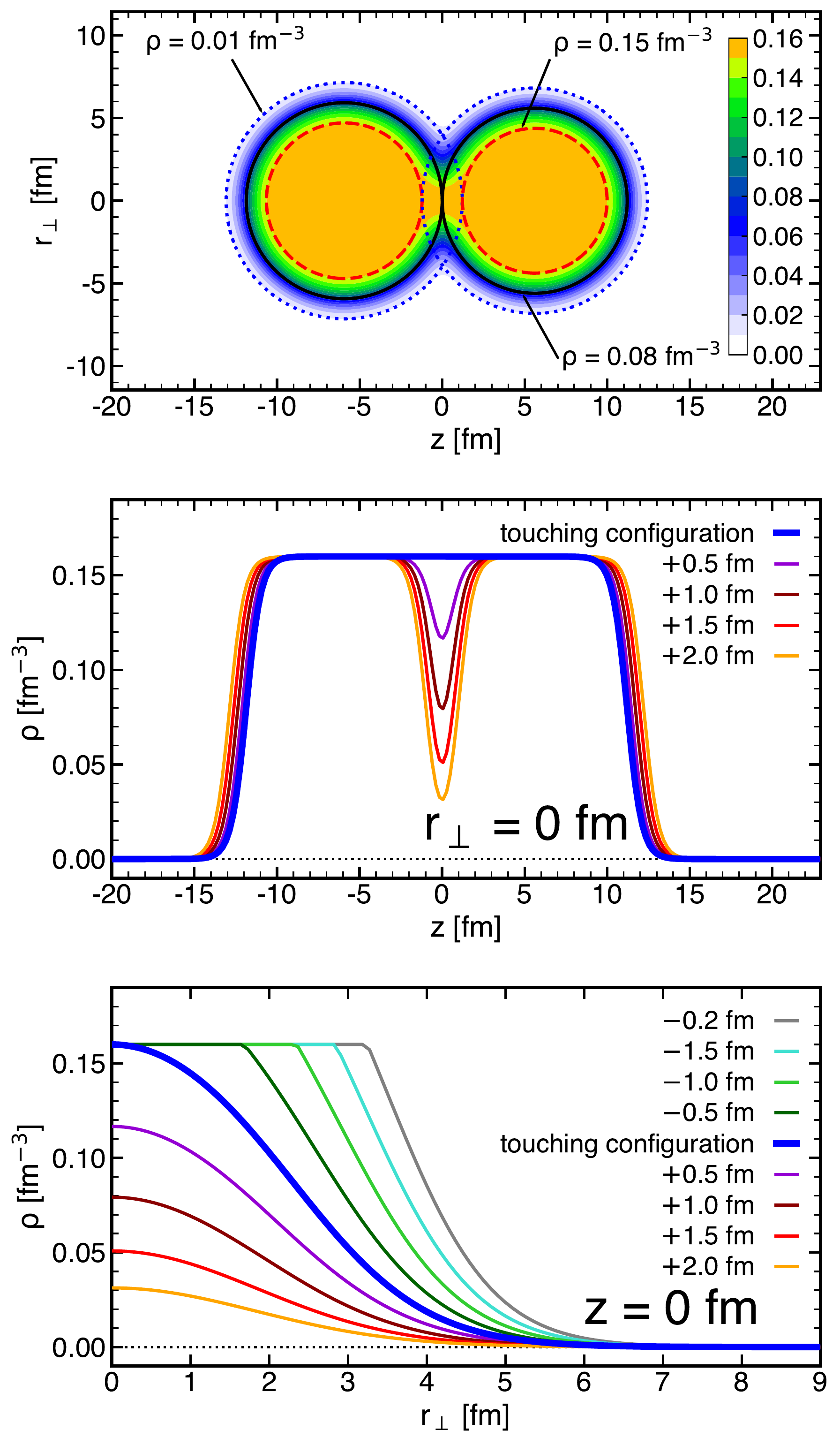}

\caption{Top: profiles of two touching spherical fragments at scission in the leptodermous model. The profiles of densities 0.01 fm$^{-3}$ and 0.15 fm$^{-3}$  are also marked. Middle: density profile of the sum of fragments' densities for the touching configuration (thick line) and for configurations where the fragments are separated by the distances indicated in the legend. Bottom: The same as in the middle panel but as a function of $r_{\perp}$.
\label{droplet}}
\end{figure}

The behavior of the nuclear density distribution in the neck region at scission can be easily understood by using a simple liquid droplet model with a leptodermous density distribution of the Fermi function type:
\begin{equation}
\rho(r) = \frac{\rho_0}{1+\exp\left(\frac{r-R_0}{a}\right)}\;. \label{fermi}
\end{equation}
For simplicity, we neglect the possible deformation of the fragments in the subsequent analysis.
In Fig. \ref{droplet} we have plotted two spherical nuclei with mass $A_L=140$ and $A_R=118$ in the touching configuration, i.e., two touching spheres, according to the liquid drop model with uniform density distribution and a sharp nucleus' surface. As mentioned in the Introduction, for leptodermous matter distribution, the touching configuration corresponds to the localization at the same point half-density radii of both fragments. Thus, in the top panel of Fig. \ref{droplet} we have marked half-density radii and also contours of densities 0.01 fm$^{-3}$ and 0.15 fm$^{-3}$ calculated for mass distribution in the form of Fermi function, Eq.  (\ref{fermi}), with $a=0.45$ fm and $\rho_0=0.16$ fm$^{-3}$. 

In the middle panel of Fig. \ref{droplet}, the density profile of this system (sum of densities of both spherical fragments) is plotted as a function of  $r_{\perp}$ and in the bottom panel as a function of $z$, similar to Fig. \ref{profiles}.
In these two panels, we have also plotted densities for configurations with fragments shifted inwards (distance between fragments decreased by $-0.5$, $-1.0$, $-1.5$, and $-2$ fm) and outwards (increased by $+0.5$, $+1.0$, $+1.5$, and $+2.0$ fm). For overlaps exceeding saturation density, we fixed the total density to be 0.16 fm$^{-3}$, despite breaking mass conservation. The obtained density profiles show a behavior similar to those calculated in the self-consistent method for asymmetric fission shown in Fig. \ref{profiles}. From the bottom panel of Fig. \ref{droplet}, it can be seen that when the fragments are pulled inwards, increasing the overlap of the fragments does not significantly change the slope of density in the direction of $r_{\perp}$. It causes an increase of the half-density radii and the size of the bulk density plateau. It is the same effect as increasing $Q_N$. However, when the fragments are pushed outwards, a reduction of the central density can be observed, similar to the results of microscopic calculation with $Q_N<4$.  

Thus, the pre-scission configuration of the self-consistent calculations corresponds to the scission density profile of the droplet model. In both cases, saturation density at the neck is obtained only at the symmetry axis. Bringing fragments closer causes an increase in neck thickness (plateau of the saturation density) and an increase of the neck parameter. Moving them away leads to a reduction of the central density and reduction of neck parameter. 

The central density in the compact mode goes down to half of the saturation density in the neck before scission. It corresponds to the fragments separated by 1 fm in the macroscopic approach.

Nonetheless, one must remember that the fragment's mass distribution is not yet fixed in the pre-scission configuration. Transfer of nucleons between the nascent fragments is still possible, although it is minimal as the rapidly decreasing energy makes the rupture of the neck almost instantaneous. The scission-point-model configuration corresponds to shapes with a density at the neck below half of the saturation density.

\section{Conclusions}

The conclusions of the paper can be summarized as follows:
\begin{itemize}
\item
The constraint on the neck parameter is a suitable tool to describe the scission configuration in self-consistent calculations.

\item
The PES before scission consists of those solutions that are stable against a reduction of the neck. In this region, decreasing the neck thickness causes an increase in energy. There is always a barrier that separates it from scission, although it may be tiny. 

\item
The pre-scission line, the edge of the fission valley, consists of the saddle points on the $Q_{30}-Q_N$ maps for fixed $Q_{20}$, which are also exit points from the valleys on the $Q_{20}-Q_N$ maps for fixed $Q_{30}$.

\item
In the pre-scission configuration of the asymmetric fission mode, the density profile of the neck in the direction perpendicular to the symmetry axis reaches the saturation density value only at $r_\perp=0$. It corresponds to a neck radius around 2 fm. 

\item 
A neck thinner than 2 fm forces a reduction of the central density. Such solutions are unstable without a constraint on the neck parameter, and the self-consistent procedure leads to a configuration with two separated fragments.

\item
In the symmetric compact mode, central density at the neck may be substantially lower and reach half of the saturation density in the pre-scission configuration.  

\item
The pre-scission configuration in the asymmetric mode corresponds to the scission configuration in the liquid droplet model, whereas in the symmetric mode, it reflects that the half-density contour of two fragments are separated by 1 fm.

\item 
The PES is continuous at scission when the neck constraint is applied. Without this constraint, the smooth surface becomes the cliff which is usually visible in the two-dimensional maps of the PES. 

\item
The calculations of the PES on the multidimensional space with the additional neck constraint show the possible pre-scission configurations at smaller quadrupole and octupole moments than those at the end of the asymmetric fission path. These fission patterns are essential for determining the fission fragments' mass distribution. 

\end{itemize}

\begin{acknowledgements}

We are grateful to Krzysztof Pomorski for the valuable discussion of the neck configuration.
R.H. and M.W. acknowledge support by the Polish National Science Center under Contract No. 2018/30/Q/ST2/00185.
The work of LMR is supported by the Spanish Ministry of Economy and Competitiveness (MINECO) Grant No. PGC2018-094583-B-I00.
\end{acknowledgements}

\bibliographystyle{apsrev}
\bibliography{pes}

\begin{thebibliography}{38}
\expandafter\ifx\csname natexlab\endcsname\relax\def\natexlab#1{#1}\fi
\expandafter\ifx\csname bibnamefont\endcsname\relax
  \def\bibnamefont#1{#1}\fi
\expandafter\ifx\csname bibfnamefont\endcsname\relax
  \def\bibfnamefont#1{#1}\fi
\expandafter\ifx\csname citenamefont\endcsname\relax
  \def\citenamefont#1{#1}\fi
\expandafter\ifx\csname url\endcsname\relax
  \def\url#1{\texttt{#1}}\fi
\expandafter\ifx\csname urlprefix\endcsname\relax\def\urlprefix{URL }\fi
\providecommand{\bibinfo}[2]{#2}
\providecommand{\eprint}[2][]{\url{#2}}

\bibitem[{\citenamefont{Krappe and Pomorski}(2012)}]{krappe2012}
\bibinfo{author}{\bibfnamefont{H.~J.} \bibnamefont{Krappe}} \bibnamefont{and}
  \bibinfo{author}{\bibfnamefont{K.}~\bibnamefont{Pomorski}},
  \emph{\bibinfo{title}{Theory of Nuclear Fission}}
  (\bibinfo{publisher}{Springer}, \bibinfo{year}{2012}).

\bibitem[{\citenamefont{Schunck and Robledo}(2016)}]{Schunck2016}
\bibinfo{author}{\bibfnamefont{N.}~\bibnamefont{Schunck}} \bibnamefont{and}
  \bibinfo{author}{\bibfnamefont{L.~M.} \bibnamefont{Robledo}},
  \bibinfo{journal}{Reports on Progress in Physics}
  \textbf{\bibinfo{volume}{79}}, \bibinfo{pages}{116301}
  (\bibinfo{year}{2016}),
  \urlprefix\url{http://stacks.iop.org/0034-4885/79/i=11/a=116301}.

\bibitem[{\citenamefont{Bender et~al.}(2020)\citenamefont{Bender, Bernard,
  Bertsch, Chiba, Dobaczewski, Dubray, Giuliani, Hagino, Lacroix, Li
  et~al.}}]{10.1088/1361-6471/abab4f}
\bibinfo{author}{\bibfnamefont{M.}~\bibnamefont{Bender}},
  \bibinfo{author}{\bibfnamefont{R.}~\bibnamefont{Bernard}},
  \bibinfo{author}{\bibfnamefont{G.}~\bibnamefont{Bertsch}},
  \bibinfo{author}{\bibfnamefont{S.}~\bibnamefont{Chiba}},
  \bibinfo{author}{\bibfnamefont{J.~J.} \bibnamefont{Dobaczewski}},
  \bibinfo{author}{\bibfnamefont{N.}~\bibnamefont{Dubray}},
  \bibinfo{author}{\bibfnamefont{S.}~\bibnamefont{Giuliani}},
  \bibinfo{author}{\bibfnamefont{K.}~\bibnamefont{Hagino}},
  \bibinfo{author}{\bibfnamefont{D.}~\bibnamefont{Lacroix}},
  \bibinfo{author}{\bibfnamefont{Z.~P.} \bibnamefont{Li}},
  \bibnamefont{et~al.}, \bibinfo{journal}{Journal of Physics G: Nuclear and
  Particle Physics}  (\bibinfo{year}{2020}),
  \urlprefix\url{http://iopscience.iop.org/10.1088/1361-6471/abab4f}.

\bibitem[{\citenamefont{Abusara et~al.}(2010)\citenamefont{Abusara, Afanasjev,
  and Ring}}]{Afanasjev2010}
\bibinfo{author}{\bibfnamefont{H.}~\bibnamefont{Abusara}},
  \bibinfo{author}{\bibfnamefont{A.~V.} \bibnamefont{Afanasjev}},
  \bibnamefont{and} \bibinfo{author}{\bibfnamefont{P.}~\bibnamefont{Ring}},
  \bibinfo{journal}{Phys. Rev. C} \textbf{\bibinfo{volume}{82}},
  \bibinfo{pages}{044303} (\bibinfo{year}{2010}),
  \urlprefix\url{https://link.aps.org/doi/10.1103/PhysRevC.82.044303}.

\bibitem[{\citenamefont{Lu et~al.}(2014)\citenamefont{Lu, Zhao, Zhao, and
  Zhou}}]{Lu2014}
\bibinfo{author}{\bibfnamefont{B.-N.} \bibnamefont{Lu}},
  \bibinfo{author}{\bibfnamefont{J.}~\bibnamefont{Zhao}},
  \bibinfo{author}{\bibfnamefont{E.-G.} \bibnamefont{Zhao}}, \bibnamefont{and}
  \bibinfo{author}{\bibfnamefont{S.-G.} \bibnamefont{Zhou}},
  \bibinfo{journal}{Phys. Rev. C} \textbf{\bibinfo{volume}{89}},
  \bibinfo{pages}{014323} (\bibinfo{year}{2014}),
  \urlprefix\url{https://link.aps.org/doi/10.1103/PhysRevC.89.014323}.

\bibitem[{\citenamefont{Rodr\'{\i}guez-Guzm\'an and
  Robledo}(2014)}]{Guzman2014}
\bibinfo{author}{\bibfnamefont{R.}~\bibnamefont{Rodr\'{\i}guez-Guzm\'an}}
  \bibnamefont{and} \bibinfo{author}{\bibfnamefont{L.~M.}
  \bibnamefont{Robledo}}, \bibinfo{journal}{Phys. Rev. C}
  \textbf{\bibinfo{volume}{89}}, \bibinfo{pages}{054310}
  (\bibinfo{year}{2014}),
  \urlprefix\url{https://link.aps.org/doi/10.1103/PhysRevC.89.054310}.

\bibitem[{\citenamefont{Rodr{\'\i}guez-Guzm{\'a}n and
  Robledo}(2016)}]{Guzman2016}
\bibinfo{author}{\bibfnamefont{R.}~\bibnamefont{Rodr{\'\i}guez-Guzm{\'a}n}}
  \bibnamefont{and} \bibinfo{author}{\bibfnamefont{L.~M.}
  \bibnamefont{Robledo}}, \bibinfo{journal}{The European Physical Journal A}
  \textbf{\bibinfo{volume}{52}}, \bibinfo{pages}{348} (\bibinfo{year}{2016}),
  \urlprefix\url{https://doi.org/10.1140/epja/i2016-16348-x}.

\bibitem[{\citenamefont{Chai et~al.}(2018)\citenamefont{Chai, Zhao, Liu, and
  Wang}}]{Chai_2018}
\bibinfo{author}{\bibfnamefont{Q.-Z.} \bibnamefont{Chai}},
  \bibinfo{author}{\bibfnamefont{W.-J.} \bibnamefont{Zhao}},
  \bibinfo{author}{\bibfnamefont{M.-L.} \bibnamefont{Liu}}, \bibnamefont{and}
  \bibinfo{author}{\bibfnamefont{H.-L.} \bibnamefont{Wang}},
  \bibinfo{journal}{Chinese Physics C} \textbf{\bibinfo{volume}{42}},
  \bibinfo{pages}{054101} (\bibinfo{year}{2018}),
  \urlprefix\url{https://doi.org/10.1088%2F1674-1137%2F42%2F5%2F054101}.

\bibitem[{\citenamefont{Warda and Zdeb}(2015)}]{Warda_2015}
\bibinfo{author}{\bibfnamefont{M.}~\bibnamefont{Warda}} \bibnamefont{and}
  \bibinfo{author}{\bibfnamefont{A.}~\bibnamefont{Zdeb}},
  \bibinfo{journal}{Physica Scripta} \textbf{\bibinfo{volume}{90}},
  \bibinfo{pages}{114003} (\bibinfo{year}{2015}),
  \urlprefix\url{https://doi.org/10.1088/0031-8949/90/11/114003}.

\bibitem[{\citenamefont{Warda et~al.}(2018)\citenamefont{Warda, Zdeb, and
  Robledo}}]{Warda18}
\bibinfo{author}{\bibfnamefont{M.}~\bibnamefont{Warda}},
  \bibinfo{author}{\bibfnamefont{A.}~\bibnamefont{Zdeb}}, \bibnamefont{and}
  \bibinfo{author}{\bibfnamefont{L.~M.} \bibnamefont{Robledo}},
  \bibinfo{journal}{Phys. Rev. C} \textbf{\bibinfo{volume}{98}},
  \bibinfo{pages}{041602(R)} (\bibinfo{year}{2018}),
  \urlprefix\url{https://link.aps.org/doi/10.1103/PhysRevC.98.041602}.

\bibitem[{\citenamefont{Zdeb et~al.}(2021)\citenamefont{Zdeb, Warda, and
  Robledo}}]{zdeb2021}
\bibinfo{author}{\bibfnamefont{A.}~\bibnamefont{Zdeb}},
  \bibinfo{author}{\bibfnamefont{M.}~\bibnamefont{Warda}}, \bibnamefont{and}
  \bibinfo{author}{\bibfnamefont{L.~M.} \bibnamefont{Robledo}},
  \bibinfo{journal}{Phys. Rev. C} \textbf{\bibinfo{volume}{104}},
  \bibinfo{pages}{014610} (\bibinfo{year}{2021}),
  \urlprefix\url{https://link.aps.org/doi/10.1103/PhysRevC.104.014610}.

\bibitem[{\citenamefont{Wilkins et~al.}(1976)\citenamefont{Wilkins, Steinberg,
  and Chasman}}]{wilkins}
\bibinfo{author}{\bibfnamefont{B.~D.} \bibnamefont{Wilkins}},
  \bibinfo{author}{\bibfnamefont{E.~P.} \bibnamefont{Steinberg}},
  \bibnamefont{and} \bibinfo{author}{\bibfnamefont{R.~R.}
  \bibnamefont{Chasman}}, \bibinfo{journal}{Phys. Rev. C}
  \textbf{\bibinfo{volume}{14}}, \bibinfo{pages}{1832} (\bibinfo{year}{1976}),
  \urlprefix\url{https://link.aps.org/doi/10.1103/PhysRevC.14.1832}.

\bibitem[{\citenamefont{Bonneau
  et~al.}(2007{\natexlab{a}})\citenamefont{Bonneau, Quentin, and
  Mikhailov}}]{Bonneau07a}
\bibinfo{author}{\bibfnamefont{L.}~\bibnamefont{Bonneau}},
  \bibinfo{author}{\bibfnamefont{P.}~\bibnamefont{Quentin}}, \bibnamefont{and}
  \bibinfo{author}{\bibfnamefont{I.~N.} \bibnamefont{Mikhailov}},
  \bibinfo{journal}{Phys. Rev. C} \textbf{\bibinfo{volume}{75}},
  \bibinfo{pages}{064313} (\bibinfo{year}{2007}{\natexlab{a}}),
  \urlprefix\url{https://link.aps.org/doi/10.1103/PhysRevC.75.064313}.

\bibitem[{\citenamefont{Bonneau
  et~al.}(2007{\natexlab{b}})\citenamefont{Bonneau, Quentin, and
  Mikhailov}}]{Bonneau07b}
\bibinfo{author}{\bibfnamefont{L.}~\bibnamefont{Bonneau}},
  \bibinfo{author}{\bibfnamefont{P.}~\bibnamefont{Quentin}}, \bibnamefont{and}
  \bibinfo{author}{\bibfnamefont{I.}~\bibnamefont{Mikhailov}}, in
  \emph{\bibinfo{booktitle}{Proceedings of International Conference on Nuclear
  Data for Science and Technology}} (\bibinfo{publisher}{EDP Sciences},
  \bibinfo{year}{2007}{\natexlab{b}}), pp. \bibinfo{pages}{343--346}.

\bibitem[{\citenamefont{Audi et~al.}(2017)\citenamefont{Audi, Kondev, Wang,
  Huang, and Naimi}}]{audi20170301}
\bibinfo{author}{\bibfnamefont{G.}~\bibnamefont{Audi}},
  \bibinfo{author}{\bibfnamefont{F.~G.} \bibnamefont{Kondev}},
  \bibinfo{author}{\bibfnamefont{M.}~\bibnamefont{Wang}},
  \bibinfo{author}{\bibfnamefont{W.~J.} \bibnamefont{Huang}}, \bibnamefont{and}
  \bibinfo{author}{\bibfnamefont{S.}~\bibnamefont{Naimi}},
  \bibinfo{journal}{Chinese Physics C} \textbf{\bibinfo{volume}{41}},
  \bibinfo{pages}{030001} (\bibinfo{year}{2017}),
  \urlprefix\url{http://hepnp.ihep.ac.cn//article/id/c7761693-0f4e-4d3c-a8da-a369a26b5823}.

\bibitem[{\citenamefont{Robledo and Bertsch}(2011)}]{PhysRevC.84.014312}
\bibinfo{author}{\bibfnamefont{L.~M.} \bibnamefont{Robledo}} \bibnamefont{and}
  \bibinfo{author}{\bibfnamefont{G.~F.} \bibnamefont{Bertsch}},
  \bibinfo{journal}{Phys. Rev. C} \textbf{\bibinfo{volume}{84}},
  \bibinfo{pages}{014312} (\bibinfo{year}{2011}),
  \urlprefix\url{https://link.aps.org/doi/10.1103/PhysRevC.84.014312}.

\bibitem[{\citenamefont{Berger et~al.}(1990)\citenamefont{Berger, Anderson,
  Bonche, and Weiss}}]{Berger90}
\bibinfo{author}{\bibfnamefont{J.~F.} \bibnamefont{Berger}},
  \bibinfo{author}{\bibfnamefont{J.~D.} \bibnamefont{Anderson}},
  \bibinfo{author}{\bibfnamefont{P.}~\bibnamefont{Bonche}}, \bibnamefont{and}
  \bibinfo{author}{\bibfnamefont{M.~S.} \bibnamefont{Weiss}},
  \bibinfo{journal}{Phys. Rev. C} \textbf{\bibinfo{volume}{41}},
  \bibinfo{pages}{R2483} (\bibinfo{year}{1990}),
  \urlprefix\url{https://link.aps.org/doi/10.1103/PhysRevC.41.R2483}.

\bibitem[{\citenamefont{P{\'e}ru and Martini}(2014)}]{Peru2014}
\bibinfo{author}{\bibfnamefont{S.}~\bibnamefont{P{\'e}ru}} \bibnamefont{and}
  \bibinfo{author}{\bibfnamefont{M.}~\bibnamefont{Martini}},
  \bibinfo{journal}{The European Physical Journal A}
  \textbf{\bibinfo{volume}{50}}, \bibinfo{pages}{88} (\bibinfo{year}{2014}),
  ISSN \bibinfo{issn}{1434-601X},
  \urlprefix\url{https://doi.org/10.1140/epja/i2014-14088-7}.

\bibitem[{\citenamefont{Robledo et~al.}(2019)\citenamefont{Robledo,
  Rodr\'{\i}guez, and Rodr\'{\i}guez-Guzm\'an}}]{Robledo2019}
\bibinfo{author}{\bibfnamefont{L.~M.} \bibnamefont{Robledo}},
  \bibinfo{author}{\bibfnamefont{T.~R.} \bibnamefont{Rodr\'{\i}guez}},
  \bibnamefont{and} \bibinfo{author}{\bibfnamefont{R.~R.}
  \bibnamefont{Rodr\'{\i}guez-Guzm\'an}}, \bibinfo{journal}{Journal of Physics
  G: Nuclear and Particle Physics} \textbf{\bibinfo{volume}{46}},
  \bibinfo{pages}{013001} (\bibinfo{year}{2019}),
  \urlprefix\url{http://stacks.iop.org/0954-3899/46/i=1/a=013001}.

\bibitem[{\citenamefont{Egido et~al.}(1997)\citenamefont{Egido, Robledo, and
  Chasman}}]{EGIDO199713}
\bibinfo{author}{\bibfnamefont{J.}~\bibnamefont{Egido}},
  \bibinfo{author}{\bibfnamefont{L.}~\bibnamefont{Robledo}}, \bibnamefont{and}
  \bibinfo{author}{\bibfnamefont{R.}~\bibnamefont{Chasman}},
  \bibinfo{journal}{Physics Letters B} \textbf{\bibinfo{volume}{393}},
  \bibinfo{pages}{13 } (\bibinfo{year}{1997}), ISSN \bibinfo{issn}{0370-2693},
  \urlprefix\url{http://www.sciencedirect.com/science/article/pii/S0370269396016024}.

\bibitem[{\citenamefont{Staszczak et~al.}(2009)\citenamefont{Staszczak, Baran,
  Dobaczewski, and Nazarewicz}}]{staszczak09}
\bibinfo{author}{\bibfnamefont{A.}~\bibnamefont{Staszczak}},
  \bibinfo{author}{\bibfnamefont{A.}~\bibnamefont{Baran}},
  \bibinfo{author}{\bibfnamefont{J.}~\bibnamefont{Dobaczewski}},
  \bibnamefont{and}
  \bibinfo{author}{\bibfnamefont{W.}~\bibnamefont{Nazarewicz}},
  \bibinfo{journal}{Phys. Rev. C} \textbf{\bibinfo{volume}{80}},
  \bibinfo{pages}{014309} (\bibinfo{year}{2009}),
  \urlprefix\url{https://link.aps.org/doi/10.1103/PhysRevC.80.014309}.

\bibitem[{\citenamefont{Staszczak et~al.}(2013)\citenamefont{Staszczak, Baran,
  and Nazarewicz}}]{Staszczak13}
\bibinfo{author}{\bibfnamefont{A.}~\bibnamefont{Staszczak}},
  \bibinfo{author}{\bibfnamefont{A.}~\bibnamefont{Baran}}, \bibnamefont{and}
  \bibinfo{author}{\bibfnamefont{W.}~\bibnamefont{Nazarewicz}},
  \bibinfo{journal}{Phys. Rev. C} \textbf{\bibinfo{volume}{87}},
  \bibinfo{pages}{024320} (\bibinfo{year}{2013}),
  \urlprefix\url{https://link.aps.org/doi/10.1103/PhysRevC.87.024320}.

\bibitem[{\citenamefont{Warda and Robledo}(2011)}]{War11}
\bibinfo{author}{\bibfnamefont{M.}~\bibnamefont{Warda}} \bibnamefont{and}
  \bibinfo{author}{\bibfnamefont{L.~M.} \bibnamefont{Robledo}},
  \bibinfo{journal}{Phys. Rev. C} \textbf{\bibinfo{volume}{84}},
  \bibinfo{pages}{044608} (\bibinfo{year}{2011}).

\bibitem[{\citenamefont{Warda et~al.}(2012)\citenamefont{Warda, Staszczak, and
  Nazarewicz}}]{wardastaszczak}
\bibinfo{author}{\bibfnamefont{M.}~\bibnamefont{Warda}},
  \bibinfo{author}{\bibfnamefont{A.}~\bibnamefont{Staszczak}},
  \bibnamefont{and}
  \bibinfo{author}{\bibfnamefont{W.}~\bibnamefont{Nazarewicz}},
  \bibinfo{journal}{Phys. Rev. C} \textbf{\bibinfo{volume}{86}},
  \bibinfo{pages}{024601} (\bibinfo{year}{2012}),
  \urlprefix\url{https://link.aps.org/doi/10.1103/PhysRevC.86.024601}.

\bibitem[{\citenamefont{Warda et~al.}(2002)\citenamefont{Warda, Egido, Robledo,
  and Pomorski}}]{War02}
\bibinfo{author}{\bibfnamefont{M.}~\bibnamefont{Warda}},
  \bibinfo{author}{\bibfnamefont{J.~L.} \bibnamefont{Egido}},
  \bibinfo{author}{\bibfnamefont{L.~M.} \bibnamefont{Robledo}},
  \bibnamefont{and} \bibinfo{author}{\bibfnamefont{K.}~\bibnamefont{Pomorski}},
  \bibinfo{journal}{Phys. Rev. C} \textbf{\bibinfo{volume}{66}},
  \bibinfo{pages}{014310} (\bibinfo{year}{2002}).

\bibitem[{\citenamefont{Schunck et~al.}(2014)\citenamefont{Schunck, Duke, Carr,
  and Knoll}}]{schunk}
\bibinfo{author}{\bibfnamefont{N.}~\bibnamefont{Schunck}},
  \bibinfo{author}{\bibfnamefont{D.}~\bibnamefont{Duke}},
  \bibinfo{author}{\bibfnamefont{H.}~\bibnamefont{Carr}}, \bibnamefont{and}
  \bibinfo{author}{\bibfnamefont{A.}~\bibnamefont{Knoll}},
  \bibinfo{journal}{Phys. Rev. C} \textbf{\bibinfo{volume}{90}},
  \bibinfo{pages}{054305} (\bibinfo{year}{2014}),
  \urlprefix\url{https://link.aps.org/doi/10.1103/PhysRevC.90.054305}.

\bibitem[{\citenamefont{Sadhukhan et~al.}(2017)\citenamefont{Sadhukhan, Zhang,
  Nazarewicz, and Schunck}}]{Sadhukhan17}
\bibinfo{author}{\bibfnamefont{J.}~\bibnamefont{Sadhukhan}},
  \bibinfo{author}{\bibfnamefont{C.}~\bibnamefont{Zhang}},
  \bibinfo{author}{\bibfnamefont{W.}~\bibnamefont{Nazarewicz}},
  \bibnamefont{and} \bibinfo{author}{\bibfnamefont{N.}~\bibnamefont{Schunck}},
  \bibinfo{journal}{Phys. Rev. C} \textbf{\bibinfo{volume}{96}},
  \bibinfo{pages}{061301(R)} (\bibinfo{year}{2017}),
  \urlprefix\url{https://link.aps.org/doi/10.1103/PhysRevC.96.061301}.

\bibitem[{\citenamefont{Verriere et~al.}(2019)\citenamefont{Verriere, Schunck,
  and Kawano}}]{Verriere19}
\bibinfo{author}{\bibfnamefont{M.}~\bibnamefont{Verriere}},
  \bibinfo{author}{\bibfnamefont{N.}~\bibnamefont{Schunck}}, \bibnamefont{and}
  \bibinfo{author}{\bibfnamefont{T.}~\bibnamefont{Kawano}},
  \bibinfo{journal}{Phys. Rev. C} \textbf{\bibinfo{volume}{100}},
  \bibinfo{pages}{024612} (\bibinfo{year}{2019}),
  \urlprefix\url{https://link.aps.org/doi/10.1103/PhysRevC.100.024612}.

\bibitem[{\citenamefont{Dubray and Regnier}(2012)}]{DUBRAY2012}
\bibinfo{author}{\bibfnamefont{N.}~\bibnamefont{Dubray}} \bibnamefont{and}
  \bibinfo{author}{\bibfnamefont{D.}~\bibnamefont{Regnier}},
  \bibinfo{journal}{Computer Physics Communications}
  \textbf{\bibinfo{volume}{183}}, \bibinfo{pages}{2035 }
  (\bibinfo{year}{2012}), ISSN \bibinfo{issn}{0010-4655},
  \urlprefix\url{http://www.sciencedirect.com/science/article/pii/S0010465512001671}.

\bibitem[{\citenamefont{Hasse and Myers}(1988)}]{hasse}
\bibinfo{author}{\bibfnamefont{R.~W.} \bibnamefont{Hasse}} \bibnamefont{and}
  \bibinfo{author}{\bibfnamefont{W.~D.} \bibnamefont{Myers}},
  \emph{\bibinfo{title}{Geometrical Relationships of Macroscopic Nuclear
  Physics}} (\bibinfo{publisher}{Springer-Verlag}, \bibinfo{year}{1988}).

\bibitem[{\citenamefont{Centelles et~al.}(2010)\citenamefont{Centelles,
  Roca-Maza, Vi\~nas, and Warda}}]{centelles10}
\bibinfo{author}{\bibfnamefont{M.}~\bibnamefont{Centelles}},
  \bibinfo{author}{\bibfnamefont{X.}~\bibnamefont{Roca-Maza}},
  \bibinfo{author}{\bibfnamefont{X.}~\bibnamefont{Vi\~nas}}, \bibnamefont{and}
  \bibinfo{author}{\bibfnamefont{M.}~\bibnamefont{Warda}},
  \bibinfo{journal}{Phys. Rev. C} \textbf{\bibinfo{volume}{82}},
  \bibinfo{pages}{054314} (\bibinfo{year}{2010}),
  \urlprefix\url{https://link.aps.org/doi/10.1103/PhysRevC.82.054314}.

\bibitem[{iae()}]{iaea}
\emph{\bibinfo{title}{Live chart of nuclides}},
  \urlprefix\url{https://www-nds.iaea.org/relnsd/vcharthtml/VChartHTML.html}.

\bibitem[{\citenamefont{Ivanyuk and Pomorski}(2009)}]{ivanyuk09}
\bibinfo{author}{\bibfnamefont{F.~A.} \bibnamefont{Ivanyuk}} \bibnamefont{and}
  \bibinfo{author}{\bibfnamefont{K.}~\bibnamefont{Pomorski}},
  \bibinfo{journal}{Phys. Rev. C} \textbf{\bibinfo{volume}{79}},
  \bibinfo{pages}{054327} (\bibinfo{year}{2009}),
  \urlprefix\url{https://link.aps.org/doi/10.1103/PhysRevC.79.054327}.

\bibitem[{\citenamefont{Ivanyuk}(2013)}]{ivanyuk12}
\bibinfo{author}{\bibfnamefont{F.}~\bibnamefont{Ivanyuk}},
  \bibinfo{journal}{Physics Procedia} \textbf{\bibinfo{volume}{47}},
  \bibinfo{pages}{17} (\bibinfo{year}{2013}), ISSN \bibinfo{issn}{1875-3892},
  \bibinfo{note}{scientific Workshop on Nuclear Fission Dynamics and the
  Emission of Prompt Neutrons and Gamma Rays, Biarritz, France, 28-30 November
  2012},
  \urlprefix\url{https://www.sciencedirect.com/science/article/pii/S1875389213004331}.

\bibitem[{\citenamefont{Randrup et~al.}(2011)\citenamefont{Randrup, M\"oller,
  and Sierk}}]{randrup11}
\bibinfo{author}{\bibfnamefont{J.}~\bibnamefont{Randrup}},
  \bibinfo{author}{\bibfnamefont{P.}~\bibnamefont{M\"oller}}, \bibnamefont{and}
  \bibinfo{author}{\bibfnamefont{A.~J.} \bibnamefont{Sierk}},
  \bibinfo{journal}{Phys. Rev. C} \textbf{\bibinfo{volume}{84}},
  \bibinfo{pages}{034613} (\bibinfo{year}{2011}),
  \urlprefix\url{https://link.aps.org/doi/10.1103/PhysRevC.84.034613}.

\bibitem[{\citenamefont{Tsekhanovich et~al.}(2019)\citenamefont{Tsekhanovich,
  Andreyev, Nishio, Denis-Petit, Hirose, Makii, Matheson, Morimoto, Morita,
  Nazarewicz et~al.}}]{Tsekhanovich2019}
\bibinfo{author}{\bibfnamefont{I.}~\bibnamefont{Tsekhanovich}},
  \bibinfo{author}{\bibfnamefont{A.}~\bibnamefont{Andreyev}},
  \bibinfo{author}{\bibfnamefont{K.}~\bibnamefont{Nishio}},
  \bibinfo{author}{\bibfnamefont{D.}~\bibnamefont{Denis-Petit}},
  \bibinfo{author}{\bibfnamefont{K.}~\bibnamefont{Hirose}},
  \bibinfo{author}{\bibfnamefont{H.}~\bibnamefont{Makii}},
  \bibinfo{author}{\bibfnamefont{Z.}~\bibnamefont{Matheson}},
  \bibinfo{author}{\bibfnamefont{K.}~\bibnamefont{Morimoto}},
  \bibinfo{author}{\bibfnamefont{K.}~\bibnamefont{Morita}},
  \bibinfo{author}{\bibfnamefont{W.}~\bibnamefont{Nazarewicz}},
  \bibnamefont{et~al.}, \bibinfo{journal}{Physics Letters B}
  \textbf{\bibinfo{volume}{790}}, \bibinfo{pages}{583 } (\bibinfo{year}{2019}),
  ISSN \bibinfo{issn}{0370-2693},
  \urlprefix\url{http://www.sciencedirect.com/science/article/pii/S0370269319300978}.

\bibitem[{\citenamefont{Ring and Schuck}(1980)}]{rin80}
\bibinfo{author}{\bibfnamefont{P.}~\bibnamefont{Ring}} \bibnamefont{and}
  \bibinfo{author}{\bibfnamefont{P.}~\bibnamefont{Schuck}},
  \emph{\bibinfo{title}{The nuclear many-body problem}}
  (\bibinfo{publisher}{Springer-Verlag}, \bibinfo{address}{New York},
  \bibinfo{year}{1980}).

\bibitem[{\citenamefont{Warda et~al.}(2014)\citenamefont{Warda, Centelles,
  Vi\~nas, and Roca-Maza}}]{warda14}
\bibinfo{author}{\bibfnamefont{M.}~\bibnamefont{Warda}},
  \bibinfo{author}{\bibfnamefont{M.}~\bibnamefont{Centelles}},
  \bibinfo{author}{\bibfnamefont{X.}~\bibnamefont{Vi\~nas}}, \bibnamefont{and}
  \bibinfo{author}{\bibfnamefont{X.}~\bibnamefont{Roca-Maza}},
  \bibinfo{journal}{Phys. Rev. C} \textbf{\bibinfo{volume}{89}},
  \bibinfo{pages}{064302} (\bibinfo{year}{2014}),
  \urlprefix\url{https://link.aps.org/doi/10.1103/PhysRevC.89.064302}.

\end{thebibliography}


\end{document}